\documentclass[fleqn,1p]{elsarticle}

\usepackage{amsmath,graphicx}

\usepackage[english]{babel}
\usepackage[latin1]{inputenc}
\usepackage[T1]{fontenc}
\usepackage{amsmath,graphicx}
\usepackage{amssymb,amsmath,epsfig,url,mathrsfs} 
\usepackage{algorithm,algorithmic}
\usepackage[usenames, dvipsnames]{color}
\usepackage[framemethod=TikZ]{mdframed}
\usepackage{xcolor}
\usepackage{listings}
\usepackage{subcaption}

\usepackage[norelsize,algo2e]{algorithm2e} %!
\newcommand*\dif{\mathop{}\!\mathrm{d}} %!
\SetKwInput{KwInput}{Input} %!

\usepackage{scalerel}

\usepackage{tikz}
\usepackage{amsfonts}
\usepackage{amsmath,amssymb}
\usepackage{systeme,mathtools}
\usetikzlibrary{positioning,arrows.meta,quotes}
\usetikzlibrary{shapes,snakes}
\usetikzlibrary{bayesnet}
\tikzset{>=latex}
\tikzstyle{plate caption} = [caption, node distance=0, inner sep=0pt, below left=0pt and 0pt of #1.south east]

\usepackage{ragged2e}
\usepackage{array}
\usepackage{booktabs}
\usepackage{siunitx}
\newcolumntype{L}[1]{>{\raggedright\let\newline\\\arraybackslash\hspace{0pt}}m{#1}}
\newcolumntype{C}[1]{>{\centering\let\newline\\\arraybackslash\hspace{0pt}}m{#1}}
\newcolumntype{R}[1]{>{\raggedleft\let\newline\\\arraybackslash\hspace{0pt}}m{#1}}

\mdfdefinestyle{MyFrame}{%
    linecolor=black,
    outerlinewidth=.5pt,
    roundcorner=5pt,
    innertopmargin=\baselineskip,
    innerbottommargin=\baselineskip,
    innerrightmargin=5pt,
    innerleftmargin=0pt,
    backgroundcolor=gray!10!white}
\definecolor{codegreen}{rgb}{0,0.6,0}
\definecolor{codegray}{rgb}{0.5,0.5,0.5}
\definecolor{codepurple}{rgb}{0.58,0,0.82}
\definecolor{backcolour}{rgb}{0.95,0.95,0.92}

\lstdefinestyle{mystyle}{
    backgroundcolor=\color{backcolour},
    commentstyle=\color{codegreen},
    keywordstyle=\color{blue},
    numberstyle=\tiny\color{codegray},
    stringstyle=\color{codepurple},
    basicstyle=\footnotesize,
    breakatwhitespace=false,
    breaklines=true,
    captionpos=b,
    keepspaces=true,
    numbers=none,
    numbersep=5pt,
    showspaces=false,
    showstringspaces=false,
    showtabs=false,
    tabsize=2
}

\usepackage{amsmath}
\usepackage{graphicx}
\usepackage[colorinlistoftodos]{todonotes}
\usepackage[colorlinks=true, allcolors=blue]{hyperref}

\graphicspath{{figures/}}

\DeclareMathAlphabet{\mathpzc}{OT1}{pzc}{m}{it}

\renewcommand{\vec}[1]{\boldsymbol{\mathrm{#1}}}

\newcommand\E{\mathbb{E}}

\renewcommand{\vec}{\boldsymbol} % Vector

\journal{Computer Speech \& Language}

\begin{document}

\begin{frontmatter}

\title{Voice Biometrics Security: Extrapolating False Alarm Rate via Hierarchical Bayesian Modeling of Speaker Verification Scores  \footnote{\copyright 2019. This manuscript version is made available under the CC-BY-NC-ND 4.0 license:\\\url{http://creativecommons.org/licenses/by-nc-nd/4.0/}}}

\author[leti,itmo,uef]{Alexey Sholokhov}
\ead{sholokhovalexey@gmail.com}

\author[uef]{Tomi Kinnunen\corref{cor1}}
\ead{tkinnu@cs.joensuu.fi}

\author[uef]{Ville Vestman\corref{cor2}}
\ead{vvestman@cs.uef.fi}

\author[nec]{Kong Aik Lee}
\ead{k-lee@ax.jp.nec.com}

\address[leti]{St. Petersburg Electrotechnical University, ul. P. Popova 5, St. Petersburg, 197376, Russia}
\address[itmo]{ITMO University, Kronverkskiy pr. 49, St. Petersburg, 197101, Russia}
\address[uef]{School of Computing, University of Eastern Finland, FI-80101, Joensuu, Finland}

\address[nec]{Biometrics Research Laboratories, NEC Corp., Tokyo, Japan}

\cortext[cor1]{Corresponding Author}

\cortext[cor2]{A part of the work of the third author was carried out during an intership at NEC.}

\begin{abstract}
How secure automatic speaker verification (ASV) technology is? More concretely, given a specific target speaker, how likely is it to find another person who gets falsely accepted as that target? This question may be addressed empirically by studying naturally confusable pairs of speakers within a large enough corpus. To this end, one might expect to find at least some speaker pairs that are indistinguishable from each other in terms of ASV. To a certain extent, such aim is mirrored in the standardized ASV evaluation benchmarks, for instance, the series of speaker recognition evaluation (SRE) organized by the National Institute of Standards and Technology (NIST). Nonetheless, arguably the number of speakers in such evaluation benchmarks represents only a small fraction of all possible human voices, making it challenging to extrapolate performance beyond a given corpus. Furthermore, the impostors used in performance evaluation are usually selected \emph{randomly}. A potentially more meaningful definition of an impostor --- at least in the context of security-driven ASV applications --- would be \emph{closest} (most confusable) other speaker to a given target.

We put forward a novel performance assessment framework to address both the inadequacy of the random-impostor evaluation model and the size limitation of evaluation corpora by addressing ASV security against closest impostors on arbitrarily large datasets. The framework allows one to make a prediction of the safety of given ASV technology, in its current state, for arbitrarily large speaker database size consisting of virtual (sampled) speakers. As a proof-of-concept, we analyze the performance of two state-of-the-art ASV systems, based on i-vector and x-vector speaker embeddings (as implemented in the popular Kaldi toolkit), on the recent VoxCeleb 1 \& 2 corpora, containing a total of 7,365 speakers. We fix the number of target speakers to 1000, and generate up to $N=100,000$ virtual impostors sampled from the generative model. The model-based false alarm rates are in a reasonable agreement with empirical false alarm rates and, as predicted, increase substantially (values up to $98\%$) with $N=100,000$ impostors. Neither the i-vector or x-vector system is immune to increased false alarm rate at increased impostor database size, as predicted by the model. 
\end{abstract}


\end{frontmatter}

%\tableofcontents

\section{Introduction}
\label{sec:intro}

Some have predicted that voice-operated user interfaces will be the next paradigm of human-machine interaction. Given that the consumer market already provides various \emph{virtual assistants} --- Google Home, Apple Siri, and Amazon Alexa to name a few --- it might be a reasonable prediction. Such services are intended to provide human-to-human like user experience leveraging from speech and speaker recognition technology, dialogue modeling and speech synthesis. An increasing number of smart services also enable users to log-in or authenticate payments using voice (or other biometric traits), for both increased security and user convenience --- there is no need to consult, \emph{e.g.}, printed key-lists (or other stealable or copiable accessories). The co-evolution of smart device technology and machine learning~\cite{Bishop-book,Goodfellow2016-deeplearning} has substantially broadened the landscape of \emph{automatic speaker verification} (ASV) \cite{Reynolds95} use cases from its traditional, highly specialized applications --- forensics and survaillance --- to our living rooms and everyday mobile environments. For instance, nowadays, smart phones, virtual assistants and other devices with powerful processors and wireless connectivity enable efficient on-device or cloud-based voice data processing, including ASV-based user authentication with algorithms that would have been difficult to execute on portable devices of the past decades. Early ASV technology, such as \cite{Reynolds95}, was developed with the aid of far less powerful computers and smaller datasets. The increase in dataset sizes and computing power has not only enabled the research community to address increasingly more challenging ASV tasks, but enabled running more powerful models in portable devices. Much of the progress in the underlying core ASV technology has been facilitated by coordinated technology benchmarks, pioneered by \emph{National Institute of Standards and Technology} (NIST) in their evaluation campaigns~\cite{Doddington2000-NIST-overview,Sadjadi2017-NIST-SRE16,Wu-2017}.  

Increased awareness of the possibilities of voice-based interaction also raises concern about the security of the technology. The possibility to invoke malicious voice commands from a distance in another user's phone \cite{Carlini2016-hidden} (potentially even using \emph{inaudible} sounds \cite{Zhang2017-dolphin}), and the potential to masquerade oneself as another targeted speaker through various \emph{spoofing attacks} \cite{Ratha2001} is widely acknowledged. The latter includes \emph{replay}, \emph{text-to-speech}, and \emph{voice conversion} attacks. Many of these \emph{technology}-aided attacks can be combated through various \emph{countermeasures} ranging from knowledge-based approaches to classification approaches, known within biometric technology standardization bodies as \emph{presentation attack detection} (PAD) \cite{isopad} methods. For instance, specialized binary detector could be used to verify liveness of a voice sample before being passed to a speaker verification system. Detection of attacks is possible since replayed speech, introduced through loudspeakers, has different frequency characteristics than live human; and since synthetic and converted voices contain processing artifacts due to training data limitations and modeling imperfections. More details of different attacks, their effectiveness, detection, and evaluation metrics are discussed elsewhere \cite{Sahidullah2019-introduction-to-PAD} in more detail. In this study, we focus on core ASV technology.

While recent efforts have capitalized the importance of preparing ASV systems against spoofing attacks, another, more fundamental question remains: how \emph{unique} the human voice is? Note that even the performance of an ASV system equipped with \emph{perfect} PAD subsystem will be upper bounded by the performance of the underlying core technology \cite{Kinnunen2018-tDCF}. This raises fundamental, yet thus far conclusively unanswered questions such as,
    \begin{itemize}
        \item Given a large-enough population of speakers (such as 7.6 billion), how likely is it to find two speakers that are confusable with each other? In other words, \emph{how many unique voices there are?}
        \item Conversely, assuming that we wish to maintain a certain minimum level of non-confusability between speakers, is there some maximum population (speaker database) size for which it can be guaranteed?
    \end{itemize}
Answers would enable both technology vendors and users of ASV technology to have increased confidence to the expected reliability of such systems. By drawing analogy from the security of passwords, some studies \cite{Nautsch2015} based on biometric information measures \cite{Lim-2016} have assessed the strength of speech representations in terms of their speaker information, though the viewpoint is rarely neither on the population size nor attacks. 

Before proceeding further, it is necessary to constrain the scope. First, the question of voice uniqueness is, clearly, ill-posed. \emph{In theory}, the number of different human voices is, if not infinite, some very large number: both the organic (physiological) and learnt traits vary greatly across individuals thanks to differences in the anatomy and kinematics of our articulatory systems --- it would be extremely unlikely to find another \emph{voice clone} with perfectly-matched voice production systems and learned traits. \emph{In practice}, when working with real-world acoustic speech waveforms, we are bounded both by \emph{extrinsic} and \emph{intrinsic} signal variations. Extrinsic variation refers to the inability to accurately measure `pure' speaker characteristics from imperfect acoustic observations (for instance, due to imperfect transducer, lossy communication channel, background noise, or reverberant environment). Intrinsic variation, in turn, refers to linguistic and non-linguistic variation induced by the speaker him/herself, some of which can be substantial \cite{Hansen2017_JasaScream,Gonzalez2017-acoustic-perceptual}. The main focus of the ASV research community for the past several decades \cite{Reynolds95,Kenny-HTPLDA} has been on improving ASV technology to handle extrinsic variations of increased complexity, though specific intrinsic factors, such as \emph{vocal effort}, have also been addressed in the context of NIST SREs \cite{NIST2010evalreport}. 

Neither the extrinsic nor intrinsic variations are deterministic, fixed operations. Therefore, there are practical limits as to how accurately one can discriminate two voices from each other. As these limits are clearly a function of the specific types of variations and distortions (as the ASV community is well aware of), it would be meaningless to attempt to answer the unconditional question of voice uniquess. The answer depends on both data conditions and the employed hypothesis tester (\emph{e.g.}, a specific human listener or a specific ASV system). We might even say that uniqueness of voices is a \emph{subjective} matter; a pair of speakers that is confusable for one hypothesis tester $A$ (for instance, a human) may not be so for another hypothesis tester $B$ (for instance, a machine).

We, therefore, constrain the focus on \emph{statistical methods} to address questions such as the above \emph{empirically} for given data. In particular, we are interested in the relation of corpus size (number of speakers) and the probability of a false alarm ($P_\text{FA}$) for a given ASV system, under a specific model detailed in Section \ref{sec:model-high-level}. The input data to our proposed model consists of detection scores (log-likelihood ratios or uncalibrated raw scores) of \emph{any} ASV system on a specific corpus. This makes the method widely applicable for the analysis of any ASV system, treated as a black-box.

The reader familiar with performance assessment of ASV systems may wonder if there is anything new to say about detection scores of a given system on a given corpus. Indeed, measuring detection errors (including $P_\text{FA}$) and calibrating speaker recognition systems is a fairly standardized activity \cite{Doddington2000-NIST-overview,Brummer2006-application-independent}. So, what is new here? The answer, in brief, is that in the NIST-style ASV evaluations, the \emph{non-target} speaker trials (pairwise comparisons of test utterances against a hypothesized speaker model with disjoint speaker identities) are, essentially, random pairs of speakers. We use more effort to model situation of more confusable (closest) pairs of speakers; one could argue a recognizer that handles the `worst cases' (closest competing) speakers well may exhibit improved generalization.

In our model, `closest' speakers are in fact \emph{none} of the non-target speakers in the training set, but \emph{virtual speakers} sampled from the distribution that models random sampling of speakers. Specifically, speakers are represented implicitly by distributions of scores corresponding to pairs of speakers. This allows us to \emph{extrapolate} beyond the given evaluation corpus to arbitrarily large virtual speaker populations. Assuming that the observed speaker pairs are sampled from a same underlying generative process, we can get an idea of how the ASV system scales up with corpus size, \emph{without collecting new speech data}.

While the technical voice conversion spoofing attacks have received a lot of attention in the recent years, it might be appropriate time to re-address worst-case impostors in the context of regular ASV as well. The initial spark for this work stems from our recent work \cite{Vestman2020-CSL-voice-mimicry} (inspired by \cite{Lau-Vulnerability2004}) where we addressed a specific research hypothesis relating to potentially emerging, yet cursorily addressed vulnerability of ASV technology \emph{against itself}. The idea was that an attacker could use (public-domain) ASV system as a voice search engine to identify suitable target speaker (specifically, the closest one), such as a celebrity or any person who uploads a lot of his/her voice or video samples to the Internet. After identifying a suitable target, the attacker would attempt to attack another ASV system (\emph{e.g.}, at bank) using natural (possibly mimicked\footnote{Mimicry is a special skill, based on the idea of a listener trying to match his or her acoustic profile with that of another person. As the acoustic correlates of speaker identity, as learned by current ASV systems, remain largely unknown, human mimicry is generally an inconsistent strategy to spoof ASV systems. This is why \cite{Vestman2020-CSL-voice-mimicry} included ASV system to first identify targets that are similar to attacker's voice.}) voice. Despite the relatively large VoxCeleb corpus with more than 7000 target speakers, none of our attackers were successful in getting falsely accepted\footnote{To be more precise, in \cite{Vestman2020-CSL-voice-mimicry}, we did not consider hard binary decisions but analyzed changes in the log-likelihood ratio (LLR) scores of the ASV systems. The nontarget LLRs, whether or not originating from zero-effort or mimicry trials were far below the range of target LLRs.}. While good news concerning security of ASV, the finding was on specific ASV systems, attackers and target corpus. One reason why the finding in \cite{Vestman2020-CSL-voice-mimicry} might have been negative is that the attacker's ASV (designed to be purposefully different from the attacked one) was not powerful enough. Nonetheless, we saw transferability across our two ASV systems in terms of relative target speaker rankings, suggesting that the attacks might be successful with a scaled-up database. We argue that there \emph{must be} a speaker database size (possibly very large) where one is likely to locate closely-matched non-target voices --- effect which we were unable to \emph{observe} under the specific experimental conditions. For these reasons, we wanted to re-address the problem by using a more principled and re-usable setup that requires neither two ASV systems (attacker's ASV and targeted ASV) nor fresh recordings. To be precise, the framework proposed in this study addresses a \emph{worst-case} attack scenario with the following two assumptions:

    \begin{enumerate}
        \item \textbf{Assumption 1: known ASV system.} The adversary's ASV system (used for identifying closest targets to attack) is the same as the attacked ASV system.
        \item \textbf{Assumption 2: access to target's enrollment data.} The adversary has access to the target speaker's enrollment data (alternatively, no domain mismatch exist between target's public-domain and enrollment recordings).
    \end{enumerate}

The generative model presented in this work enables us to increase the corpus size indefinitely to establish empirical performance bounds on the false alarm rate, under these two assumptions. As search queries to the attacked system can be limited and the enrollment utterances can be protected by template protection techniques, neither assumption is necessarily realistic from the perspective of the adversary. An evaluation corpus designer, technology vendor, or a bank, however, may still want to assess worst-case performance. Importantly, the above assumptions greatly simplify the set-up over the scenarios addressed in \cite{Vestman2020-CSL-voice-mimicry}. The methods developed in this study can be seen as an extension of the arsenal of statistical performance evaluation tools. We address each of the two assumptions in the empirical part.

We summarize our two main contributions as follows. First, we propose a general-purpose performance metric, \emph{worst-case false alarm rate with $N$ impostors} ($P_\text{FA}^N$). It is the probability of accepting the closest impostor among $N$ available candidate impostors selected randomly for each enrolled speaker. As will be discussed below, the proposed metric reduces to the `conventional' probability of a false alarm ($P_\text{FA}$) if $N=1$. Second, we devise a hierarchical Bayesian generative model of non-target score distribution to enable prediction of $\mathrm{P}_\text{FA}^N$ for arbitrarily large values of $N$ that can exceed the number of non-target speakers in a given corpus. The proposed model allows one to make a prediction of the safety of given ASV technology, in its current state, for arbitrarily large speaker database size consisting of virtual (sampled) speakers.  Importantly, as the training data consists of detection scores only, the framework is widely applicable for the analysis of arbitrary ASV system (or even other biometric systems). Further, all the model parameters are automatically inferred from data, leaving no manually-tunable control parameters to be set. As a representative snapshot of the current ASV technology and evaluation databases, our proof-of-concept experiments include two widely-used ASV methods based on \emph{i-vector} \cite{Dehak2011-front-end-FA} and \emph{x-vector} \cite{SnyderGSPK18} embeddings, evaluated on the combined VoxCeleb1 \cite{nagrani2017voxceleb} and VoxCeleb2 \cite{Chung18b} corpora.

\section{Measuring and Extrapolating False Alarm Rates}\label{sec:model-high-level}

An automatic speaker verification (ASV) system is a hypothesis testing machine that takes a pair of speech utterances $\mathcal{X}=(\mathcal{X}_\text{e},\mathcal{X}_\text{t})$ --- one for enrollment, one for test --- and produces a numerical detection score $s \in \mathbb{R}$, with the convention that higher values (in relative terms) indicate stronger support for the \emph{same speaker} (null) hypothesis and low scores for the \emph{different speaker} (alternative) hypothesis. Speech utterances are typically represented as fixed-sized \emph{speaker embeddings} such as \emph{i-vectors} \cite{Dehak2011-front-end-FA} or \emph{x-vectors} \cite{SnyderGSPK18} and the detection score is a \emph{logarithmic likelihood ratio} (LLR) produced by a statistical back-end model, such as the \emph{probabilistic linear discriminant analysis} (PLDA) \cite{Prince-LIV,Kenny-HTPLDA}. 

\subsection{False alarm rate}

The detection score $s$ can be interpreted as a realization of a continuous random variable that admits an underlying probability density $p(s)$, with $p(s)\geq 0$ and $\int_{s=-\infty}^\infty p(s) \dif{s}=1$. In the conventional ASV set-up (as in NIST SREs \cite{Doddington2000-NIST-overview,Sadjadi2017-NIST-SRE16}), the performance of an ASV system is assessed using two types of users, \emph{targets} and \emph{nontargets}. The former means that speaker identities of $\mathcal{X}_\text{e}$ and $\mathcal{X}_\text{t}$ match, while the latter means that they differ. We denote the class-conditional score densities of targets and nontargets by $p(s|\text{tar})$ and $p(s|\text{non})$, respectively. 

Our focus is on ASV security against impostors, characterized by the nontarget score distribution. In specific, an ASV system is characterized by the probability of accepting a random impostor (sometimes known as \emph{zero-effort} impostor), known as \emph{false alarm rate} (or \emph{false acceptance rate}). It is defined as the following non-increasing function of detection  threshold $\tau \in \mathbb{R}$, 
\begin{equation}
\mathrm{P}_\text{FA}(\tau) = \int_\tau^{\infty} p(s|\text{non}) \dif s,
\end{equation}
where $\tau$ is fixed in advance to set $P_\text{FA}(\tau)$ to a desirable level (increasing $\tau$ reduces false alarm rate but increases target rejection rate, also known as \emph{miss rate}). 

As we do not have access to $p(s|\text{non})$, in practice $\mathrm{P}_\text{FA}(\tau)$ is usually approximated using \emph{Monte-Carlo} (MC) methods \cite{Robert-MC}. Monte-Carlo integration is a class of numerical methods that can be used to evaluate expected values of complicated functions. It replaces integrals in expectations by finite sums with the help of independent samples drawn from the underlying probability distribution. By using $\mathbb{I}\{\cdot\}$ to denote an indicator function that equals 1 for a true proposition and 0 otherwise, we write the MC-approximated false alarm rate as,  
\begin{equation} \label{eq:fa}
\begin{aligned}
\mathrm{P}_\text{FA}(\tau) & = \int_{-\infty}^{\infty} p(s|\text{non}) \mathbb{I}\{s > \tau\} \dif s\\ 
& = \E_{s \sim p(s|\text{non})}[\mathbb{I}\{s > \tau\}] \approx \frac{1}{R} \sum_{r=1}^R \mathbb{I}\{s_r > \tau\}, \;\; s_r \sim p(s|\text{non}),
\end{aligned}
\end{equation}
by assuming one is able to obtain $R$ independent samples $s_r$ from the non-target score distribution. Here, $\E_{s \sim p(s)}[g(s)]$ denotes expected (average) value of function $g(s)$ w.r.t. the distribution $p(s)$. Usually we have just a finite collection of detection scores $\{s_r\}_{r=1}^R$ with no further knowledge of $p(s|\text{non})$.

\subsection{Reinterpreting False Alarm Rate as Averaged Speaker-Pair Conditioned False Alarm Rate}

In the following, we provide an alternative view of the false alarm rate as an average of speaker-pair specific false alarm rates, useful in paving way towards a new performance metric and a generative model designed to extrapolate its values beyond available datasets. To that end, note first that the detection scores $\{s_r\}$ are obtained through an ASV system that processes some pre-defined \emph{trial list} formed from a finite set of pairwise speaker comparisons. Thus, the terms in  \eqref{eq:fa} can be divided into groups corresponding to unique pairs of speakers. In the special case when these groups are of equal size, we can rewrite the sum in \eqref{eq:fa} as
\begin{equation} \label{eq:fa-ssum}
\frac{1}{R} \sum_{r=1}^R \mathbb{I}\{s_r > \tau\} = \frac{1}{T} \sum_{i=1}^T \underbrace{ \frac{1}{L_i} \sum_{l=1}^{L_i} \mathbb{I}\{s_{i,l} > \tau\} }_{\text{speaker-pair specific}\atop\text{probability of a false alarm}},
\end{equation}
where $s_{i,l}$ denotes the $l^\text{th}$ trial score from speaker pair $i$, $L_i$ is the total number of scores for the $i^\text{th}$ speaker pair, and $T$ is the total number of speaker pairs, such that $L_1=L_2=...=L_T=L$ and $R=T \cdot L$. Here, the inner sum can be interpreted as the probability of a trial from a given pair of speakers being incorrectly accepted, with the outer sum forming average of the speaker-pair specific false alarm probabilities.

\begin{figure}[!t]
\centering
\includegraphics[width=13cm]{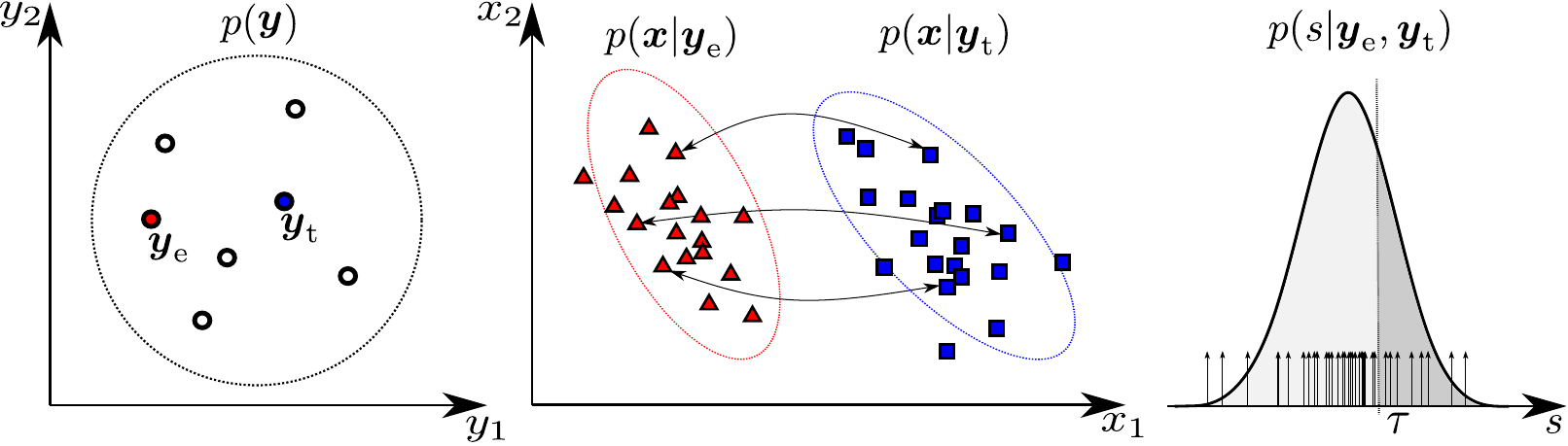}
\caption{Illustration of the steps to obtain the speaker-pair conditioned score distribution. (Left) Latent speaker identity space. Each element of this space corresponds to unique identity. Small circles represent a dataset consisting of 7 speakers. (Middle) Observation space. Here, $p(\vec{x}|\vec{y})$ is the distribution of utterances $\vec{x}$ of the speaker $\vec{y}$. (Right) Score space. Here, $p(s|\vec{y}_\text{e}, \vec{y}_\text{t})$ is the distribution of similarity scores between utterances of the pair of speakers. Samples from this distribution are shown as vertical arrows. Shaded area corresponds to the speaker-pair conditioned false alarm probability which depends on the decision threshold $\tau$.}
\label{fig:spk-fa}
\end{figure}

The above simple reformulation provides a bridge towards our proposed framework detailed below. As our approach enables extrapolation of $P_\text{FA}(\tau)$ estimates beyond a given speech corpus, it is necessary to proceed from the empirical averaged false alarm rate \eqref{eq:fa-ssum} towards a continuous-space formulation. In specific, as illustrated in Fig. \ref{fig:spk-fa}, we require a model that enables sampling both speakers and speaker-pair specific scores from continuous distributions. Note, first, that the distribution of non-target scores $p(s|\text{non})$ can be seen as a continuous mixture of score distributions between all possible pairs of speakers,
\begin{equation} \label{eq:fa-mixture}
p(s|\text{non}) = \iint p(s|\vec{y}_\text{e}, \vec{y}_\text{t}) p(\vec{y}_\text{e}) p(\vec{y}_\text{t}) \dif \vec{y}_\text{e} \dif \vec{y}_\text{t},
\end{equation}
where we have introduced two new vector-valued variables $\vec{y}_\text{e}$ and $\vec{y}_\text{t}$, viewed as so-called \emph{latent identity variables} \cite{Prince-LIV, Kenny-HTPLDA}. Let $\vec{y} \in \mathcal{Y}$ be an element of some space $\mathcal{Y}$. The latent identity variable framework \cite{Prince-LIV} assumes that $\vec{y}$ is a pure representation of a person's identity and that there is a distribution on $\mathcal{Y}$ with known probability density function $p(\vec{y})$. Given a likelihood function for the latent identity variable (\emph{e.g.}, \emph{meta-embedding} \cite{Brummer-2018}), one can make inferences about speaker identities within a set of speech utterances. Examples of such tasks include speaker verification, identification and clustering \cite{Brummer-2010}. For instance, speaker verification involves testing whether two sets of utterances belong to the same or to different speakers. In this setup the unit of observations, a speech utterance, corresponds to a single speaker identity. 

The same framework can also be used in the score domain where observations correspond to pairs of identities. Given a pair of (unknown) identity variables $\vec{y}_\text{e}, \vec{y}_\text{t} \in \mathcal{Y}$, one can describe the distribution of similarity scores between the corresponding speakers by the density function $p(s|\vec{y}_\text{e}, \vec{y}_\text{t})$. This allows to conduct a test of alternative hypotheses such as: (i) two sets of scores belong to different pairs of speakers, (ii) two sets of scores share one common speaker, (iii) two sets of scores belong to the same pair of speakers. 

The representation \eqref{eq:fa-mixture} allows us to rewrite the false alarm probability  $\mathrm{P}_\text{FA}(\tau)$ akin to \eqref{eq:fa-ssum}, namely, 
\begin{gather} \nonumber
\mathrm{P}_\text{FA}(\tau) = \int_\tau^{\infty} p(s|\text{non}) \dif s = \int_\tau^{\infty} \left( \iint p(s|\vec{y}_\text{e}, \vec{y}_\text{t}) p(\vec{y}_\text{e}) p(\vec{y}_\text{t}) \dif \vec{y}_\text{e} \dif \vec{y}_\text{t} \right) \dif s \\
= \iint \underbrace{\left( \int_\tau^{\infty} p(s|\vec{y}_\text{e}, \vec{y}_\text{t}) \dif s \right)}_{\text{speaker-pair specific}\atop\text{probability of a false alarm}}  p(\vec{y}_\text{e}) p(\vec{y}_\text{t}) \dif \vec{y}_\text{e} \dif \vec{y}_\text{t},\label{eq:rewritten-fa-rate}
\end{gather}
where the inner integral is the speaker-pair specific probability of a false alarm and the outer two integrals correspond to summing over all possible speaker pairs.

Given a trial list with speaker IDs, one can obtain the estimate of $\mathrm{P}_\text{FA}(\tau)$ using so-called \emph{nested} Monte-Carlo \cite{Rainforth-2018}. It uses MC estimate of the inner integral in \eqref{eq:rewritten-fa-rate} to compute MC estimate of the outer integral. The corresponding nested sampling scheme consists of sampling a pair of speakers, followed by sampling a set of scores from the speaker-pair specific score distribution. In practice, any trial list consisting of $T$ unique speaker pairs and the corresponding scores can be thought as being generated according to this scheme. For instance (see Figure \ref{fig:spk-fa}), the following generative process produces the scores suitable for computing the nested MC estimate of $\mathrm{P}_\text{FA}(\tau)$:
\begin{enumerate}
  \item sample an enrolled speaker $\vec{y}_\text{e}^{(i)} \sim p(\vec{y})$
  \item sample a test speaker $\vec{y}_\text{t}^{(i)} \sim p(\vec{y})$  
  \item sample $n_\text{e}$ utterances of the enrolled speaker $\vec{x}_{e,j} \sim p(\vec{x}|\vec{y}_\text{e}^{(i)}),j=1,2,\dots,n_\text{e}$
  \item sample $n_\text{t}$ utterances of the test speaker $\vec{x}_{\text{t},k} \sim p(\vec{x}|\vec{y}_\text{t}^{(i)}),k=1,2,\dots,n_\text{t}$  
  \item compute $L_i = n_\text{e} \cdot n_\text{t}$ pairwise scores $s_{j,k} = \mathrm{score}(\vec{x}_{e,j}, \vec{x}_{\text{t},k})$ using an ASV system.
\end{enumerate}
Here, the index $i$ runs over all speaker pairs and $p(\vec{x}|\vec{y})$ denotes the conditional distribution of speech utterances $\vec{x}$ belonging to speaker $\vec{y}$. Here, the last step can be equivalently re-formulated as sampling from the distribution of scores conditioned on a pair of speakers:
\begin{enumerate}
  \item sample an enrolled speaker $\vec{y}_\text{e}^{(i)} \sim p(\vec{y})$
  \item sample a test speaker $\vec{y}_\text{t}^{(i)} \sim p(\vec{y})$   
  \item sample $L_i$ scores $s_{l} \sim p(s|\vec{y}_\text{e}^{(i)}, \vec{y}_\text{t}^{(i)}), l=1,2,\dots,L_i$.
\end{enumerate}

Now, the nested MC estimate of \eqref{eq:rewritten-fa-rate} can be found as
\begin{gather}  \label{eq:fa-as-average}
\mathrm{P}_\text{FA}(\tau) \approx \frac{1}{T} \sum_{i=1}^T \mathrm{P}_\text{FA}^{(i)}(\tau),
\end{gather}
where
\begin{gather} \nonumber
\mathrm{P}_\text{FA}^{(i)}(\tau) = \int_\tau^{\infty} p(s|\vec{y}_\text{e}^{(i)}, \vec{y}_\text{t}^{(i)}) \dif s \\ \approx \frac{1}{L_i} \sum_{l=1}^{L_i} \mathbb{I}\{s_l > \tau\}, \;\; s_l \sim p(s|\vec{y}_\text{e}^{(i)},  \vec{y}_\text{t}^{(i)}), \;\; \vec{y}_\text{e}^{(i)}, \vec{y}_\text{t}^{(i)}  \sim p(\vec{y}).
\end{gather}
We refer to $\mathrm{P}_\text{FA}^{(i)}(\tau)$ as \emph{speaker-pair conditioned} false alarm rate. It is the fraction of similarity scores between these speakers being above the decision threshold $\tau$. 

The $\mathrm{P}_\text{FA}(\tau)$ can be estimated based on either a model or available empirical data.
In the former case one needs a probabilistic model of between-speaker similarity scores and an algorithm to generate samples from this model. In specific, one must be able to obtain samples from the distribution of speaker identities $p(\vec{y})$ and from the distribution of similarity scores $p(s|\vec{y}_\text{e}, \vec{y}_\text{t})$ given an arbitrary speaker pair $(\vec{y}_\text{e}, \vec{y}_\text{t})$. An example of such a model will be described in Section \ref{sec:generative-model}. In the latter case, the distribution $p(\vec{y})$ is a uniform distribution over speakers' IDs and the observed between-speaker scores can be viewed as being samples drawn from an unknown distribution $p(s|\vec{y}_\text{e}, \vec{y}_\text{t})$. That is, the $\mathrm{P}_\text{FA}(\tau)$ can be estimated by repeated selection of random pairs of speakers from a dataset and computing similarity scores between random subsets of their sessions. Algorithm \ref{algo:fa-zero-effort} summarizes a procedure to estimate the probability of accepting a zero-effort impostor, $\mathrm{P}_\text{FA}(\tau)$, given a set of utterances with speaker labels.

\begin{algorithm}[!h] 
\setlength{\abovedisplayskip}{1pt}
\setlength{\belowdisplayskip}{1pt}
%\setstretch{0.5}
\SetAlgoLined
\KwInput{Dataset with speaker labels}
\KwResult{$\mathrm{P}_\text{FA}(\tau)$}
 \For{$i = 1...T$}
 {
  $\,$ Select random enrolled (target) speaker, $\vec{y}_\text{e}^{(i)}$\\
  Select random test speaker, $\vec{y}_\text{t}^{(i)}$\\
  Compute $L_i$ scores $\{s_l\}$ between $\vec{y}_\text{e}^{(i)}$ and $\vec{y}_\text{t}^{(i)}$ \\
  Compute the MC estimate of the speaker-pair conditioned false alarm probability:
  \begin{gather} \nonumber
  \mathrm{P}_\text{FA}^{(i)}(\tau) \approx \frac{1}{L_i} \sum_{l=1}^{L_i} \mathbb{I}\{s_l > \tau\},
  \end{gather}  
 }
 Compute the MC estimate of $\mathrm{P}_\text{FA}(\tau)$:
  \begin{gather} \nonumber
  \mathrm{P}_\text{FA}(\tau) \approx \frac{1}{T} \sum_{i=1}^T \mathrm{P}_\text{FA}^{(i)}(\tau)
  \end{gather}
 \caption{} \label{algo:fa-zero-effort}
\end{algorithm}

One should note that in general case, \emph{i.e.}, when speaker-pair specific subsets have different number of scores, $L_i$, the estimators defined by \eqref{eq:fa} and \eqref{eq:fa-as-average} produce different results. The former estimator relies on the unrealistic i.i.d. assumption and does not take into account data dependencies resulting from multiple appearances of the same speaker in a given trial list. In practice, however, limited resources usually do not allow to collect sufficiently many unique pairs of speakers to satisfy this assumption. As a result, the estimate may be biased if some speaker pairs have disproportionately large number of trials compared to the rest. The estimator in \eqref{eq:fa-as-average} compensates this bias by assigning weights to the terms in the sum which are inversely proportional to the number of trials. A more in-depth discussion of data dependence in speaker recognition evaluation can be found in \cite{Wu-2017}.

\subsection{Worst-Case False Alarm Rate With $N$ Impostors}

As \eqref{eq:fa-as-average} suggests, the probability of accepting an impostor speaker can be estimated by averaging the speaker-pair conditioned false alarm probabilities. In particular, Algorithm~\ref{algo:fa-zero-effort} repeats simulation of the zero-effort attack scenario where an impostor speaker is selected at random from the general population. 

\begin{figure}[!t]
\centering
\includegraphics[width=6cm]{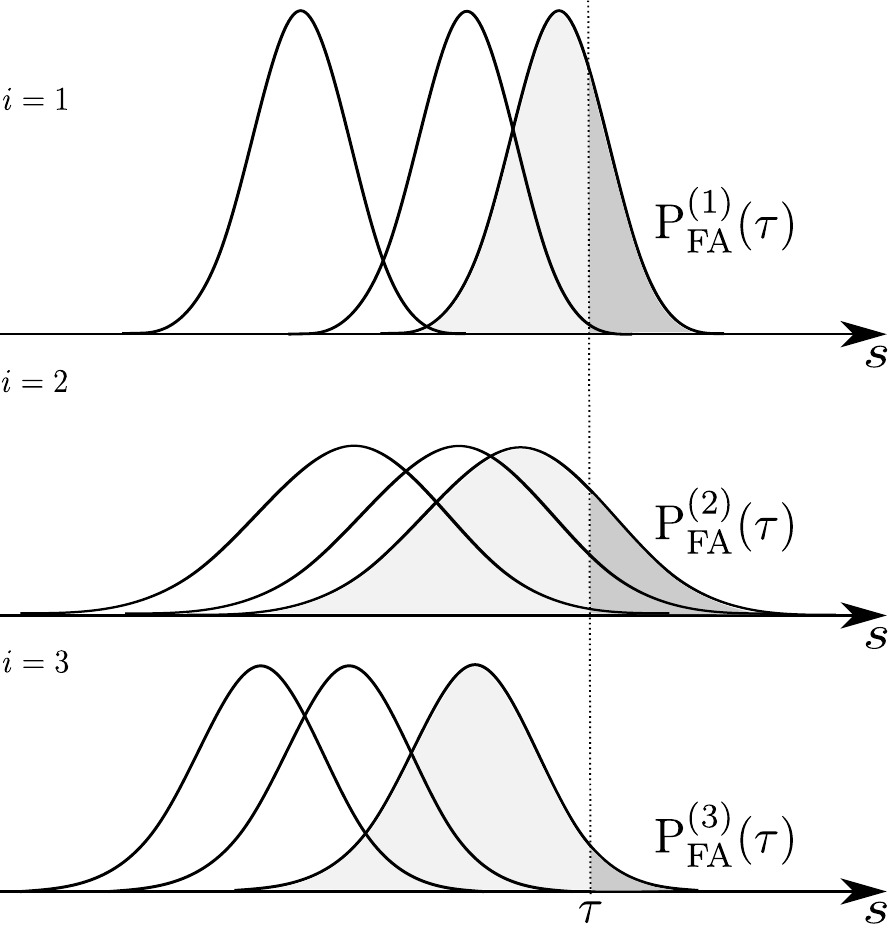}
\caption{Illustration of a few iterations of the Algorithm \ref{algo:fa-n-best} in the case of $N=3$. At each iteration the algorithm samples $N$ non-target speakers. The corresponding speaker-pair conditioned score distributions are depicted as Gaussians. The algorithm selects a distribution with the largest mean value. This distribution is shown as the one with shaded area under the curve. Finally, the algorithm computes the probability of the score being above the decision threshold $\tau$ for the selected distribution. This probability, denoted as $\mathrm{P}_{\mathrm{FA}}^{(i)}$, equals the area under the curve to the right of $\tau$. Since the score distributions are not available in practice, one can only compute the empirical estimates of $\mathrm{P}_{\mathrm{FA}}^{(i)}$.}
\label{fig:scores}
\end{figure}

We propose a new characteristic of ASV systems which generalizes $\mathrm{P}_\text{FA}(\tau)$ to attack scenarios where an impostor speaker is selected among $N$ speakers with the intention to fool an ASV system. We call it the \emph{worst-case false alarm rate with $N$ impostors}, denoted by $\mathrm{P}^N_\text{FA}(\tau)$. Algorithm \ref{algo:fa-n-best} outlines the steps to estimate $\mathrm{P}^N_\text{FA}(\tau)$. Here, $\text{similarity} (\cdot, \cdot)$ is an arbitrary similarity measure between speakers. The similarity function could be defined, for instance, in a speaker embedding space. In this work, all our models are defined in the score domain. One possible strategy to select the closest speaker is to sample $N$ sets of scores from $p(s|\vec{y}_\text{e}^{(i)}, \vec{y}_{\text{t},j}^{(i)})$ for $j=1, \dots,N$ and select the set with the highest mean value. We adopt this strategy. Figure \ref{fig:scores} illustrates progression of Algorithm \ref{algo:fa-n-best}.

\begin{algorithm}[!h] 
\setlength{\abovedisplayskip}{1pt}
\setlength{\belowdisplayskip}{1pt}
%\setstretch{0.5}
\SetAlgoLined
\KwInput{Dataset with speaker labels}
\KwResult{$\mathrm{P}^N_\text{FA}(\tau)$}
 \For{$i = 1...T$}
 {
  $\,$ Select random enrolled (target) speaker, $\vec{y}_\text{e}^{(i)}$\\
  Select $N$ random test speakers, $\vec{y}_{\text{t},1}^{(i)}, \vec{y}_{\text{t},2}^{(i)},...,\vec{y}_{\text{t},N}^{(i)}$\\
  Find the closest speaker $\vec{y}_{\text{t},k}^{(i)}$, where
  \begin{gather} \nonumber
  k = \arg\max_j \text{similarity} (\vec{y}_{e}^{(i)}, \vec{y}_{\text{t},j}^{(i)})
  \end{gather}
  Compute $L_i$ scores $\{s_l\}$ between $\vec{y}_\text{e}^{(i)}$ and $\vec{y}_{\text{t},k}^{(i)}$ \\
  Compute the MC estimate of the speaker-pair conditioned false alarm probability:
  \begin{gather} \nonumber
  \mathrm{P}_\text{FA}^{(i)}(\tau) \approx \frac{1}{L_i} \sum_{l=1}^{L_i} \mathbb{I}\{s_l > \tau\},
  \end{gather}  
 }
Compute the MC estimate of $\mathrm{P}^N_\text{FA}(\tau)$:
\begin{gather} \nonumber
\mathrm{P}^N_\text{FA}(\tau) \approx \frac{1}{T} \sum_{i=1}^T \mathrm{P}_\text{FA}^{(i)}(\tau)
\end{gather}
\caption{} \label{algo:fa-n-best}
\end{algorithm}

\begin{figure}[!h]
\centering
\includegraphics[width=13cm]{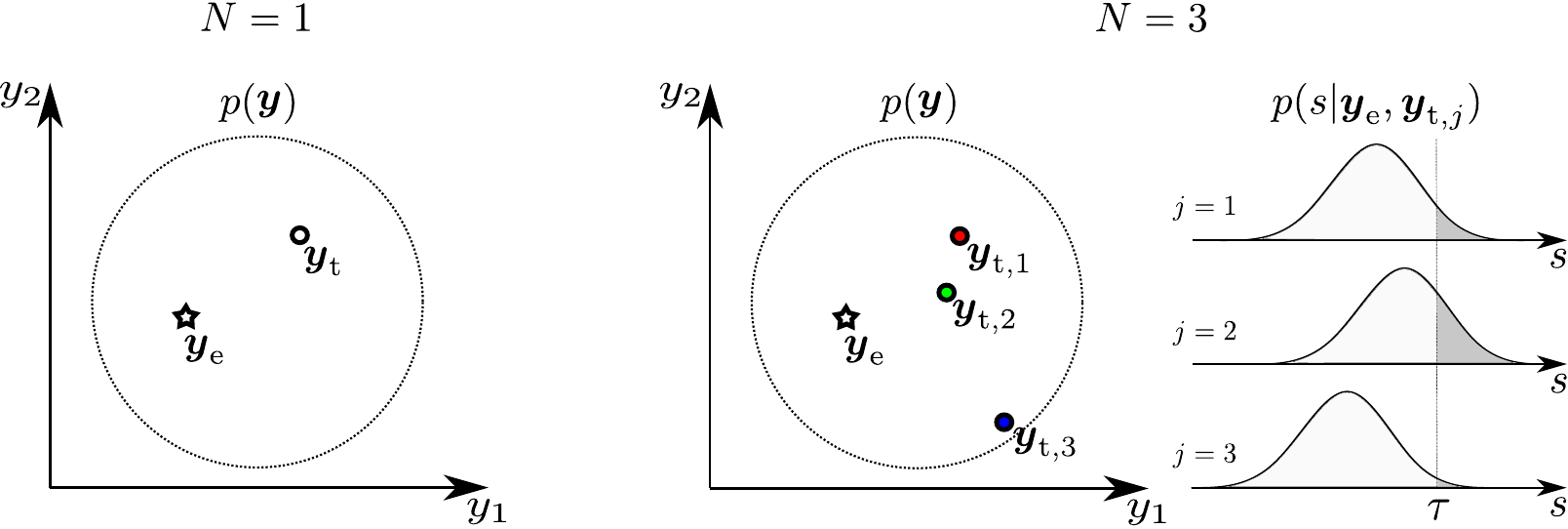}
\caption{Illustration of two evaluation scenarios where an impostor speaker is selected among $N=1$ (zero-effort attack) and $N=3$ impostor speakers. (Left and Middle) Latent speaker identity space. Star represents an enrolled speaker and circles correspond to the impostor speakers. (Right) Distributions of scores between the enrolled speaker $\vec{y}_\text{e}$ and each of the $N$ impostor speakers $\vec{y}_{\text{t},j}$ for $j=1,\dots,N$.}
\label{fig:pfa-n}
\end{figure}

Algorithm \ref{algo:fa-n-best} reduces to the zero-effort imposture case if $N=1$, or if one selects a \emph{random} (among $N$ available) test speaker, rather than the closest one to the enrolled speaker. 
Figure \ref{fig:pfa-n} demonstrates differences between these cases.

\subsection{Performance Extrapolation Through Generative Model of Scores} \label{sec:generative-model}

Note that in the above strategy, the value of $N$ is limited by the number of speakers in the dataset. Here, we describe an approach to extrapolate $\mathrm{P}^N_\text{FA}(\tau)$ for values of $N$ greater than the number of speakers in a dataset. Our main assumption is 
to approximate the speaker-pair conditioned score distribution $p(s|\vec{y}_\text{e}, \vec{y}_\text{t})$ as a (univariate) Gaussian. It should be noted that this assumption, by itself, does not put too many constraints on the shape of the distribution $p(s|\mathrm{non})$ which can be asymmetric and/or heavy-tailed.

In the sequel we describe a probabilistic model of between-speaker scores which follows the generative process in Algorithm \ref{algo:fa-n-best}. It will allow to obtain estimates of $\mathrm{P}^N_\text{FA}(\tau)$ for arbitrary values of $N$. We introduce two sets of latent variables: $\vec{\eta}_\text{e}$ and $\vec{\eta}_\text{t}$. The variables $\vec{\eta}_\text{e}$ are \emph{shared} among $N$ speaker pairs and represent individual characteristics of the enrolled speaker $\vec{y}_\text{e}$. The variables $\vec{\eta}_{\text{t},j}$, in turn, are responsible for differences between score distributions within a set of test speakers $\vec{y}_{\text{t},j}$.

The proposed probabilistic model consists of the distribution of observations $p(s|\vec{\eta}_\text{e}, \vec{\eta}_\text{t})$, which is assumed to be Gaussian, and the prior distribution of latent variables $p(\vec{\eta}_\text{e}, \vec{\eta}_\text{t}) = p(\vec{\eta}_\text{t} | \vec{\eta}_\text{e})p(\vec{\eta}_\text{e})$. Assuming that one can generate random samples of these variables, sampling scores from the model can be done according to the following steps (index $i$ is omitted for clarity):
\begin{enumerate}
  \item sample $\vec{\eta}_\text{e} \sim p(\vec{\eta}_\text{e})$
  \item sample $\vec{\eta}_{\text{t},j} \sim p(\vec{\eta}_\text{t}|\vec{\eta}_\text{e})$ for $j=1...N$ 
  \item sample $N$ sets of scores $\mathcal{S}_j=\{s_{j,l}\}$, where $s_{j,l} \sim p(s|\vec{\eta}_\text{e}, \vec{\eta}_{\text{t},j})$
\end{enumerate}
We consider a particular instance of such model where $\vec{\eta}_\text{t} = \{\mu\}$, $\vec{\eta}_\text{e} = \{m, \lambda, \sigma^2\}$ and the joint probability density function of the observed score and latent variables is factorized as follows
\begin{gather} \nonumber
p(s, \vec{\eta}_\text{e}, \vec{\eta}_\text{t}) = p(s|\mu, \sigma^2)p(\mu|m, \lambda, \sigma^2)p(m)p(\lambda)p(\sigma^2).
\end{gather}
The individual factors are outlined below:
\begin{gather} \nonumber
p(s|\mu, \sigma^2) = \mathcal{N}(s|\mu, \sigma^2) \\ \nonumber
p(\mu|m, \lambda, \sigma^2) = \mathcal{N}(\mu|m, \sigma^2/\lambda) \\ \nonumber
p(m) = \mathcal{N}(m|\mu_0, \sigma^2_0) \\ \nonumber
p(\lambda) = \mathrm{Gam}(\lambda|\alpha_\lambda, \beta_\lambda) \\ \nonumber
p(\sigma^2) = \mathrm{InvGam}(\sigma^2|a_\sigma, b_\sigma). \nonumber
\end{gather}
Here, $\vec{\theta} = \{\mu_0, \sigma_0^2, a_\sigma, b_\sigma\, \alpha_\lambda, \beta_\lambda\}$ are hyper-parameters which can be estimated on the training set of scores formed according to Algorithm \ref{algo:fa-n-best}. Given hyper-parameters, the model can be used to predict $\mathrm{P}^N_\text{FA}(\tau)$ for arbitrary values of $N$ using Algorithm \ref{algo:fa-n-best-model}. It differs from Algorithm \ref{algo:fa-n-best} in a way that the observed scores are replaced by samples from a generative model meant to approximate the unknown distribution of scores. In the special case of the proposed model the $\mathrm{P}^N_\text{FA}(\tau)$ can be estimated without explicit sampling of scores. The assumption of the speaker-pair conditioned score distribution being Gaussian allows to compute the estimate as
\begin{gather} \nonumber
\mathrm{P}^N_\text{FA}(\tau) \approx \frac{1}{T} \sum_{i=1}^T 1 - \Phi(\tau|\max_{j=1...N}(\{\mu_{i,j}\}), \sigma^2_i),
\end{gather}
where $m_i$, $\lambda_i$, $\sigma^2_i$ and $\mu_{i,j}$ are sampled from the corresponding distributions. Here, $\Phi(\cdot)$ denotes cumulative distribution function of the Gaussian distribution.

\begin{algorithm}[!h] 
\setlength{\abovedisplayskip}{1pt}
\setlength{\belowdisplayskip}{1pt}
%\setstretch{0.5}
\SetAlgoLined
\KwInput{Generative model of scores}
\KwResult{$\mathrm{P}^N_\text{FA}(\tau)$}
 \For{$i = 1...T$}
 {
  $\,$ Sample $N$ sets of scores from the model: $\mathcal{S}_j$ for $j=1...N$ \\
  Find the set with the highest mean score
  \begin{gather} \nonumber
  k = \arg\max_j \text{mean}(\mathcal{S}_j)
  \end{gather}
  Compute the MC estimate of the speaker-pair conditioned false alarm probability:
  \begin{gather} \nonumber
  \mathrm{P}_\text{FA}^{(i)}(\tau) \approx \frac{1}{|\mathcal{S}_k|} \sum_{l=1}^{|\mathcal{S}_k|} \mathbb{I}\{s_l > \tau\}, s_l \in \mathcal{S}_k
  \end{gather}  
 }
Compute the MC estimate of $\mathrm{P}^N_\text{FA}(\tau)$:
\begin{gather} \nonumber
\mathrm{P}^N_\text{FA}(\tau) \approx \frac{1}{T} \sum_{i=1}^T \mathrm{P}_\text{FA}^{(i)}(\tau)
\end{gather}
 \caption{} \label{algo:fa-n-best-model}
\end{algorithm}

This model assumes shared variance among score distributions $p(s|\vec{y}_\text{e}, \vec{y}_{\text{t},j})$ for $j=1,\dots,N$ given a target speaker $\vec{y}_\text{e}$. This assumption as well as the choice of specific distributions are primarily motivated by the convenience of computing the posterior distribution of the latent variables.
In particular, using \emph{conjugate pairs} of distributions \cite{Gelman-book} as building blocks in the model allows to devise efficient algorithms to obtain approximate posterior distribution. This leads to closed-form updates in the \emph{expectation-maximization} (EM) algorithm \cite{Dempster-1977} used to estimate the model hyper-parameters, with the details provided in Appendix I. Further insight to the form of the score distributions implied by our model (including its limitations) is provided in Appendix II. In Section \ref{subsec:discussion} we provide discussion of the adequacy of the model assumptions and potential alternatives.

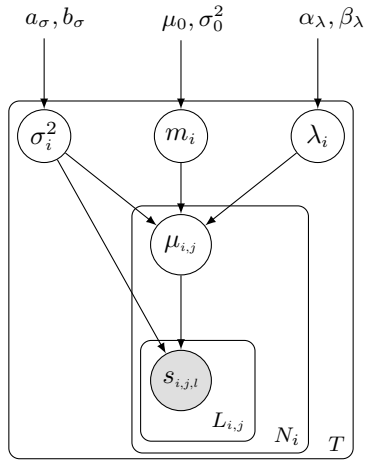
\begin{figure}[!h]
\centering
\begin{tikzpicture}[scale=1.2]

    \node [obs,minimum size=0.8cm] (s) at (0,0) {$s_{\scaleto{i,j,l}{4pt}}$};

    \node [text width=0.01cm] (fake1) at (0.6, -0.1) {};  
    \plate [color=black] {part1} {(s)(fake1)} {$L_{\scaleto{i,j}{4pt}}$};
    
    \node [circle,draw=black,fill=white,inner sep=0pt,minimum size=0.8cm] (mu) at (0,1.5) {$\mu_{\scaleto{i,j}{4pt}}$};
 
    \node [text width=0.01cm] (fake2) at (1.2, -0.4) {};  
    \plate [inner sep=0.1cm, color=black] {part2} {(s)(mu)(fake2)(part1.west)} {$N_{\scaleto{i}{4pt}}$};
    \path [draw,->] (mu) edge (s);
    
    \node [circle,draw=black,fill=white,inner sep=0pt,minimum size=0.7cm] (m) at (0,2.7) {$m_{\scaleto{i}{4pt}}$};
    \path [draw,->] (m) edge (mu);
    \node [circle,draw=black,fill=white,inner sep=0pt,minimum size=0.7cm] (lambda) at (1.5,2.7) {$\lambda_{\scaleto{i}{4pt}}$};    
    \path [draw,->] (lambda) edge (mu);    
    \node [circle,draw=black,fill=white,inner sep=0pt,minimum size=0.7cm] (sigma) at (-1.5,2.7) {$\sigma^2_{\scaleto{i}{4pt}}$};    
    \path [draw,->] (sigma) edge (mu);     
    \path [draw,->] (sigma) edge (s);   
    \node [text width=0.1cm] (fake3) at (-1.5, -0.5) {};  
    \plate [inner sep=0.1cm, color=black] {part3} {(m)(lambda)(sigma)(fake3)} {$T$};  
    
    \node [text width=0.5cm] (m_params) at (0,4.) {\small $\mu_0,\sigma^2_0$};
    \path [draw,->] (m_params) edge (m);  

    \node [text width=0.5cm] (lambda_params) at (1.5,4.) {\small $\alpha_\lambda, \beta_\lambda$};
    \path [draw,->] (lambda_params) edge (lambda);  
    
    \node [text width=0.5cm] (sigma_params) at (-1.5,4.) {\small $a_\sigma, b_\sigma $};
    \path [draw,->] (sigma_params) edge (sigma);      

\end{tikzpicture}
\caption{A graphical representation of the generative model. Here, $T$ is the number of target speakers, $N_i$ is the number of non-target speakers for the $i^\text{th}$ target speaker, and $L_{i,j}$ is the number of similarity scores between the $i^\text{th}$ target speaker and the $j^\text{th}$ non-target speaker.} \label{fig:bayes-net}
\end{figure}

Figure \ref{fig:bayes-net} depicts the Bayesian network of the proposed model. A Bayesian network is a directed graphical model \cite{Bishop-book} that represents a set of random variables and their conditional dependencies via a directed acyclic graph. Empty circles denote latent variables, shaded circles denote observed variables and nodes without circles denote deterministic parameters. A group of nodes surrounded by a box, called a plate, labeled with $T$ indicates that the subgraph inside a plate is duplicated $T$ times \cite{Buntine-1994}. The arrows between the nodes point from the parent variables to their children variables and represent the conditional dependencies between these variables.

\section{Experimental Setup}

This section describes the ASV systems, protocols, and the dataset we use for the experiments with the proposed worst-case false alarm rate with $N$ impostors ($\mathrm{P}^N_\text{FA}$) metric.

\subsection{Dataset}

A suitable dataset for our experiments has to fulfill two requirements. First, it must have a large number of speakers to not only train well-performing ASV systems, but to  have enough speakers in the evaluation side to produce good $\mathrm{P}^N_\text{FA}$ estimates from the ASV scores. Second, each speaker in the evaluation side should have enough utterances to produce a sufficiently large number of scores between each pair of speakers, required for reliable $\mathrm{P}^N_\text{FA}$ estimation. For these reasons, we chose the VoxCeleb datasets (VoxCeleb1 \cite{nagrani2017voxceleb} \& VoxCeleb2 \cite{Chung18b}). When combined, the datasets contain $7365$ speakers and, on average, each speaker has well over 100 utterances, which typically originate from about 20 sessions.

We divided the available speakers into three disjoint sets containing $5345$, $40$, and $2000$ speakers. The first set of $5345$ speakers is used to train the ASV systems. The second set of $40$ speakers consists of the test speakers in the standard VoxCeleb1 ASV evaluation protocol, which is used for evaluating performance of our ASV systems. The third, gender-balanced set contains $1000$ male and $1000$ female speakers and is used for the experiments with $\mathrm{P}^N_\text{FA}$ estimation. The speakers in this last set were chosen so that each had utterances from at least 18 different sessions; otherwise the split between the first and the last set was random.

\subsection{Automatic Speaker Verification Systems}

\renewcommand\arraystretch{1.3}
\begin{table}[h]
\caption{Details of the ASV systems used in this study.} 
\centering \small
\begin{tabular}{l>{\raggedright}p{0.33\linewidth}>{\raggedright\arraybackslash}p{0.33\linewidth}}
\toprule
& i-vector system & x-vector system \\
\midrule
Acoustic features & $24$-dimensional MFCCs + delta + double-delta coefficients; energy based speech activity detection &  $30$-dimensional MFCCs; energy based speech activity detection \\
Background model & Gaussian mixture model of $2048$ components with full covariance matrices; trained using the whole training data &  --- \\
Embedding extractor & Trained with $\num[group-separator={\text{\,}}]{100000}$ longest utterances in the training set & Trained using the whole training data plus $\num[group-separator={\text{\,}}]{1000000}$ utterances obtained by data augmentation (reverb, noise, babble, music) \\
Embeddings & $400$-dimensional i-vectors & $512$-dimensional x-vectors \\
LDA and PLDA & Both trained using the whole training data; dimensionality reduction to $200$-D with LDA & Both trained using the whole training data; dimensionality reduction to $200$-D with LDA \\
\bottomrule
\end{tabular}
\label{tab:asv-systems}
\end{table}
\renewcommand\arraystretch{1.0}

We provide experimental results for two different ASV systems, based on the two most commonly used speaker embeddings, i-vectors \cite{Dehak2011-front-end-FA} and x-vectors \cite{SnyderGSPK18}. We trained both systems using Kaldi \cite{povey2011kaldi} recipes for VoxCeleb using our custom train-test data division. Both systems use mel-frequency cepstral coefficients (MFCCs) as acoustic features and a combination of \emph{linear discriminant analysis} (LDA) and \emph{probabilistic} LDA (PLDA) in the scoring backend. The fundamental difference between the i-vector and x-vector systems is that the former is based on Gaussian generative model, while the latter is trained discriminatively and utilizes longer time context via time-delay neural network. Another major difference is that the x-vector system is trained with a larger training set leveraging from data augmentation. For further details of the systems, refer to Table \ref{tab:asv-systems}. 

\subsection{Evaluation Protocols}
We used two ASV protocols to serve two different purposes. First, we adopted the standard VoxCeleb1 ASV protocol to assess the performance of our ASV systems. This protocol contains $40$ speakers, and $37720$ evaluation trials with a balanced number of target (same speaker) and non-target (different speaker) trials. The second protocol is used to obtain a large number of non-target scores for a large number of speaker pairs to estimate $\mathrm{P}^N_\text{FA}$. For each of the $2000$ speakers in the testing set, we randomly chose $18$ utterances so that all the utterances were from different sessions. Then, for each pair of speakers, we obtained $18^2 = 324$ trials by forming all the utterance pairs between the two speakers. In total, we had $\num[group-separator={\text{\,}}]{1999000}\cdot 324 = \num[group-separator={\text{\,}}]{647676000}$ trials, where $\num[group-separator={\text{\,}}]{1999000}$\footnote{$2000!/(2!(2000-2)!) = \num[group-separator={\text{\,}}]{1999000}$ (number of 2-combinations in a set of 2000 speakers)} is the total number of unique speaker pairs. The above number includes cross-gender trials. Including only speaker pairs within one gender, we have $\num[group-separator={\text{\,}}]{161838000}$ trials for both males and females.

\section{Results}

\subsection{Performance of Speaker Verification Systems}

\renewcommand\arraystretch{1.1}
\begin{table}[h]
\caption{Parameters of three different detection cost functions (DCF) used in this study. The system thresholds~($\tau$) that mimimize these DCFs are used to estimate false alarm rates in the following experiments.}
\centering \small
\begin{tabular}{l C{2cm} C{2cm} C{2cm}}
\toprule
& $P_\textrm{target}$ & $C_\textrm{miss}$ & $C_\textrm{fa}$ \\
\midrule
minDCF$_1$ & 0.5 & 10 & 1\\
minDCF$_2$ & 0.5 & 1 & 1\\
minDCF$_3$ & 0.5 & 1 & 10\\
\bottomrule
\end{tabular}
\label{tab:dcf_parameters}
\end{table}
\renewcommand\arraystretch{1.0}

Before proceeding to our proposed generative approach, we validate correctness of the ASV implementations through standard performance metrics. To this end, we report \emph{equal error rate} (EER) and \emph{minimum normalized detection cost function} (minDCF) \cite{alvin2004nist}. EER is obtained by setting the system threshold $\tau$ so that false alarm and miss rates equal each other. 
The threshold selection for minDCF, in turn, is governed by the parameters $P_\textrm{target}$ (prior probability of target speaker), $C_\textrm{miss}$ (cost of missing the target speaker), and $C_\textrm{fa}$ (cost of falsely accepting a non-target speaker). For this study, we adopt three different sets of parameters (Table \ref{tab:dcf_parameters}): the first set has high cost for misses, the second set has equal costs for misses and false alarms, and the last one penalizes false alarms more. From the security perspective, DCF$_3$ is the most relevant, whereas the other two DCFs can be utilized in applications where high security is not required.

\renewcommand\arraystretch{1.1}
\begin{table}[b!]
\caption{Performance of i-vector and x-vector systems on VoxCeleb1 test protocol. The original protocol (`all') contains both intra- and inter-gender trials. The numbers under the category `pooled' are computed using only intra-gender trials from both genders.}
\centering \small
\begin{tabular}{L{2cm} C{2cm} C{2cm} C{2cm} C{2cm}}
\toprule
& minDCF$_1$ & minDCF$_2$ & minDCF$_3$ & EER (\%)\\
\midrule
\textbf{i-vector} & & & & \\
\quad male & 0.43 & 0.14 & 0.31 & 6.97\\
\quad female & 0.53 & 0.17 & 0.37 & 8.80\\
\quad pooled & 0.44 & 0.14 & 0.34 & 7.18\\
\quad all & 0.30 & 0.11 & 0.27 & 5.62\\
\textbf{x-vector} & & & & \\
\quad male & 0.27 & 0.09 & 0.24 & 4.71\\
\quad female & 0.30 & 0.10 & 0.24 & 5.19\\
\quad pooled & 0.28 & 0.09 & 0.23 & 4.75\\
\quad all & 0.21 & 0.07 & 0.19 & 3.61\\
\bottomrule
\end{tabular}
\label{tab:asv-performance}
\end{table}
\renewcommand\arraystretch{1.0}

Table \ref{tab:asv-performance} shows the EERs and minDCFs for i-vector and x-vector systems using VoxCeleb test protocol (category `all'). In addition, we split the protocol based on genders and also report a result for the `pooled' category, which does not contain inter-gender trials making the original test protocol more difficult. Our results are in line with the results reported in the original Kaldi recipes. As we used about $2000$ speakers less for system training, our EER for x-vector system is about $0.5$\% (absolute) higher than what is reported in the original recipe.

In addition to the overall performance difference between i-vector and x-vector systems, these systems differ in their ability to recognize speakers from different genders. For the i-vector system, the performance for males is considerably better, whereas for the x-vector system the difference between the genders is smaller.

\begin{figure}[h]
 
  \begin{subfigure}[b]{0.5\textwidth}
    \includegraphics[width=\textwidth]{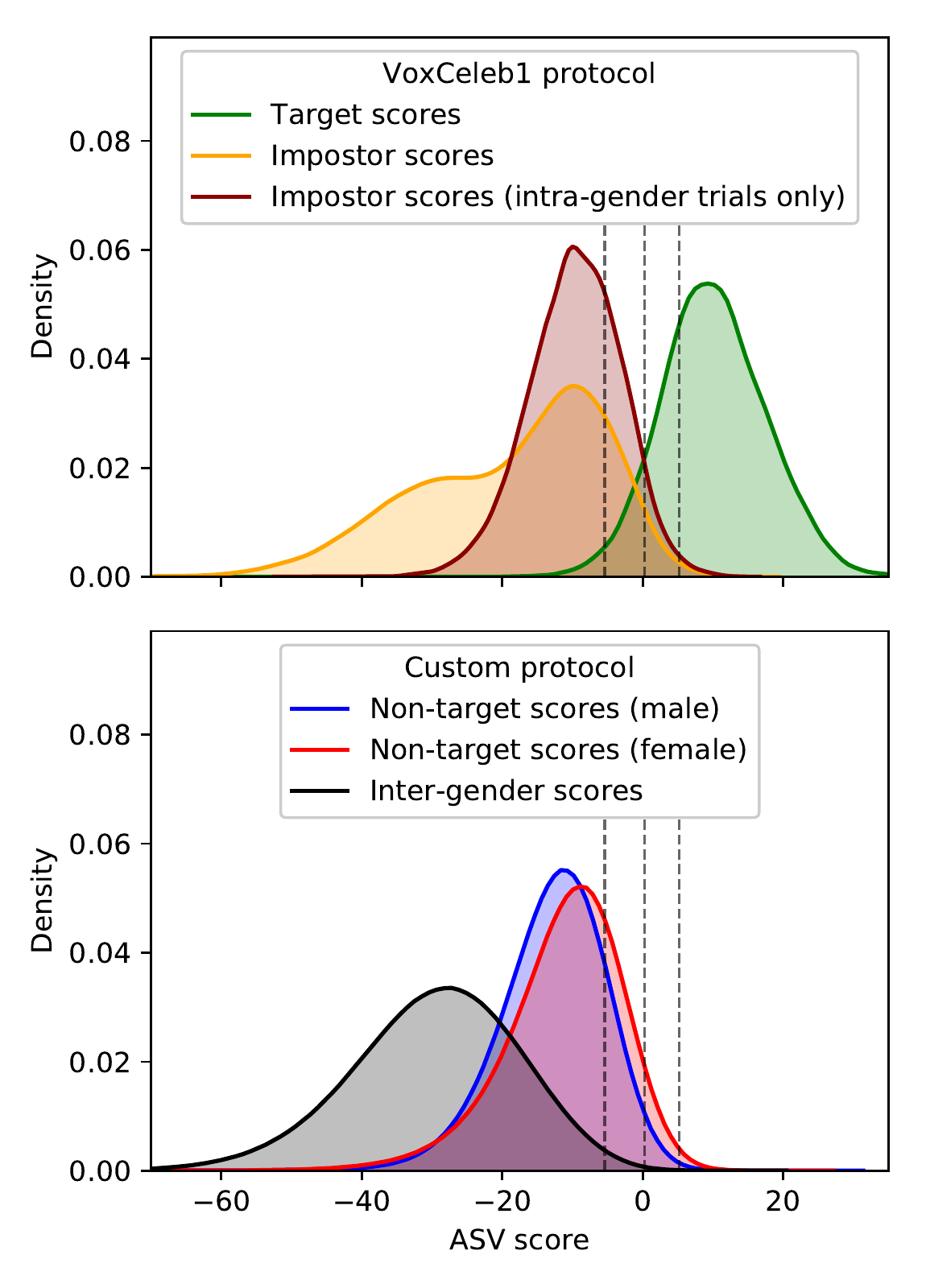}
    \caption{i-vector.}
    \label{fig:f1}
  \end{subfigure}
  \hfill
  \begin{subfigure}[b]{0.5\textwidth}
    \includegraphics[width=\textwidth]{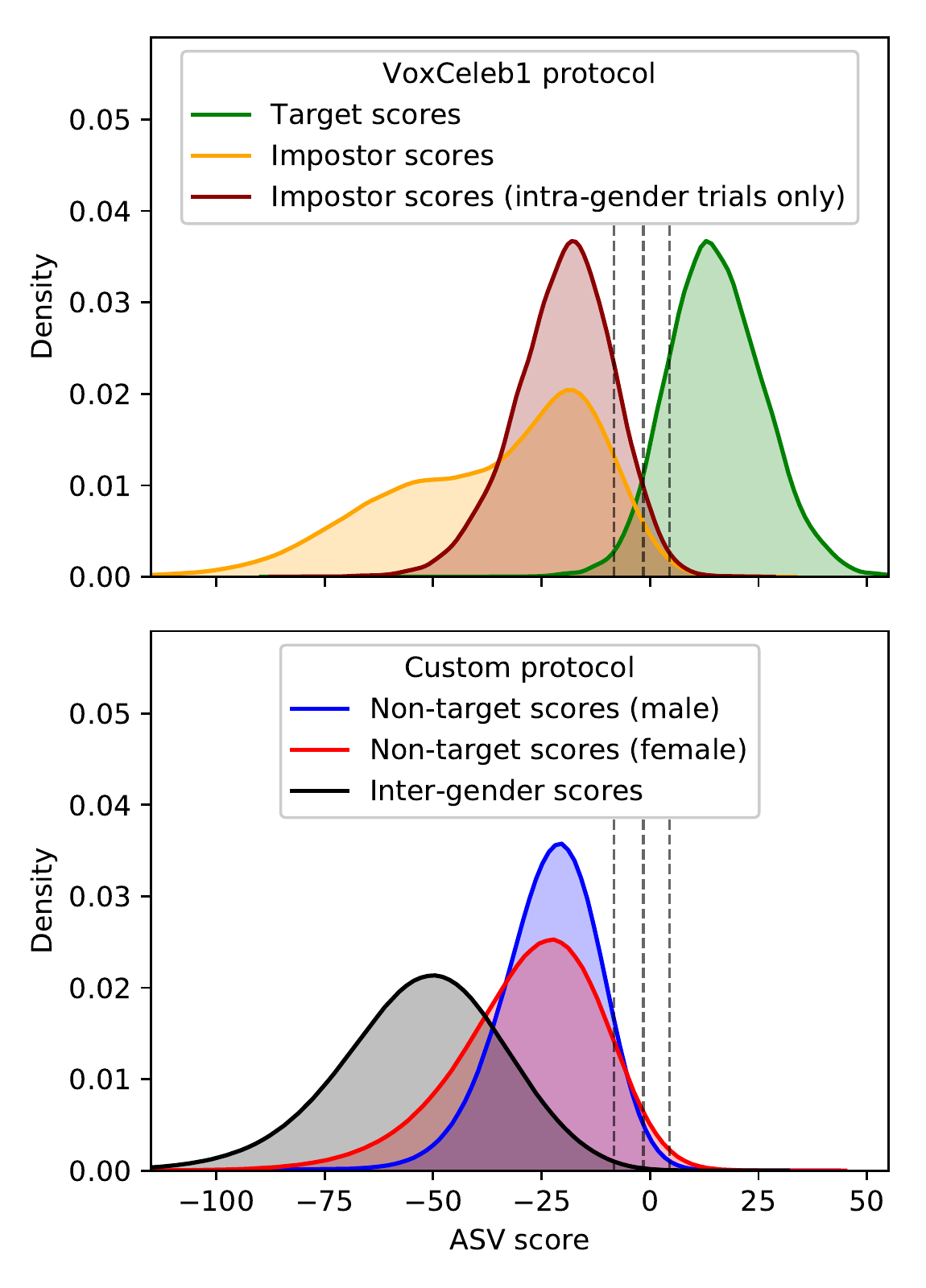}
    \caption{x-vector.}
    \label{fig:f2}
  \end{subfigure}
  \caption{Score distributions for the standard VoxCeleb protocol and the custom protocol obtained using i-vector and x-vector systems. Dashed lines represent minDCF thresholds for `pooled' scores (see Table \ref{tab:asv-performance}) using different sets of cost parameters presented in Table \ref{tab:asv-performance}. \label{fig:asv_score_distributions}}
\end{figure}

In Figure \ref{fig:asv_score_distributions}, we display score distributions for the VoxCeleb1 test protocol and for our custom protocol containing non-target scores only. The VoxCeleb1 protocol is used to set the system thresholds $\tau$ for the $\mathrm{P}^N_\text{FA}$ estimation experiments presented in the next section. In these experiments, we use gender-specific thresholds obtained via minimizing DCF separately on male and female trials. Note that for clarity, Figure~\ref{fig:asv_score_distributions} does not show gender-specific thresholds, but instead it shows the thresholds for `pooled' category.

\subsection{Estimation of Worst-Case False Alarm Rates}

\begin{figure}[h!]
  \begin{subfigure}[b]{0.5\textwidth}
    \includegraphics[width=\textwidth]{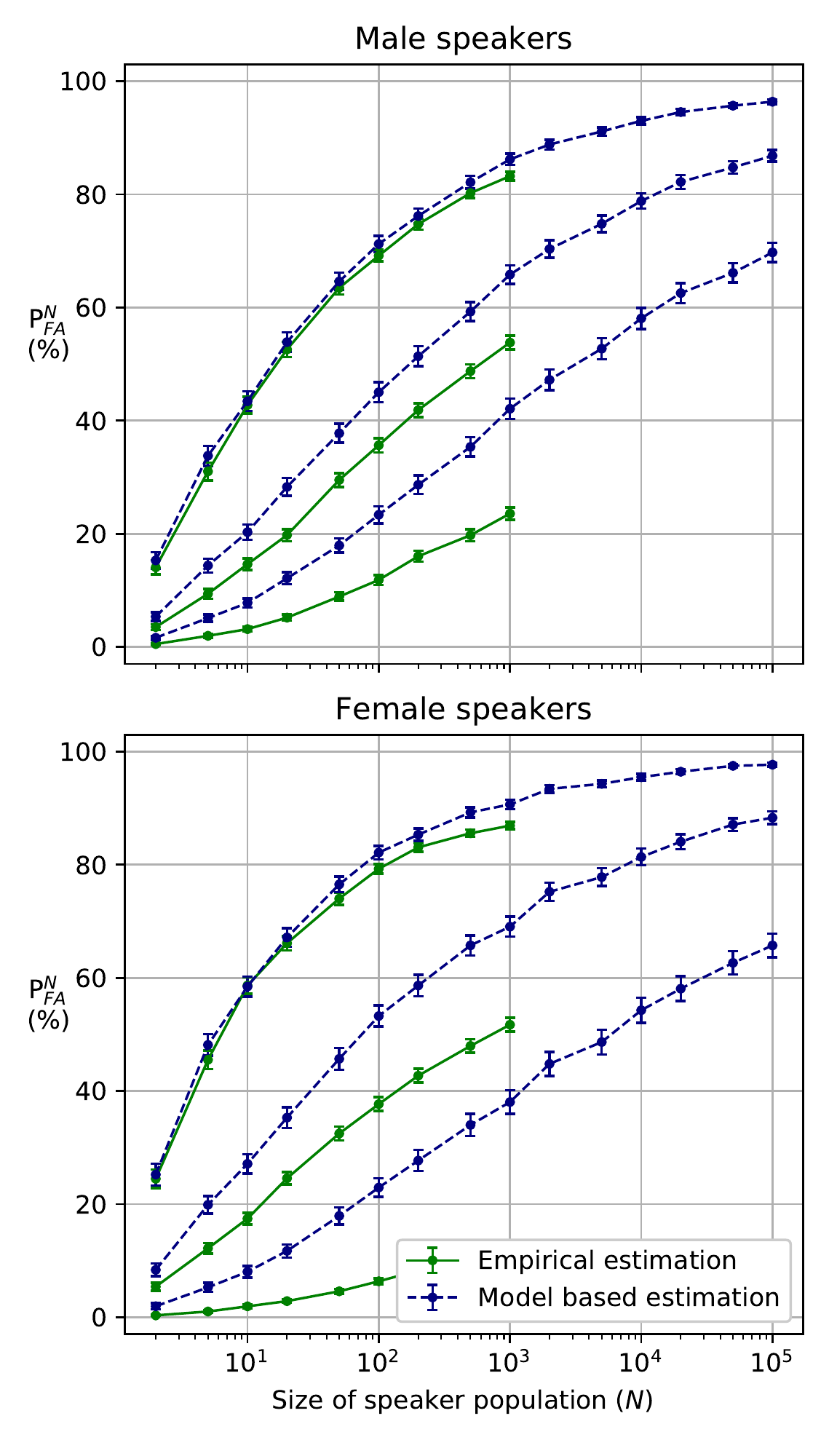}
    \caption{i-vector.}
    \label{fig:f1_2}
  \end{subfigure}
  \hfill
  \begin{subfigure}[b]{0.5\textwidth}
    \includegraphics[width=\textwidth]{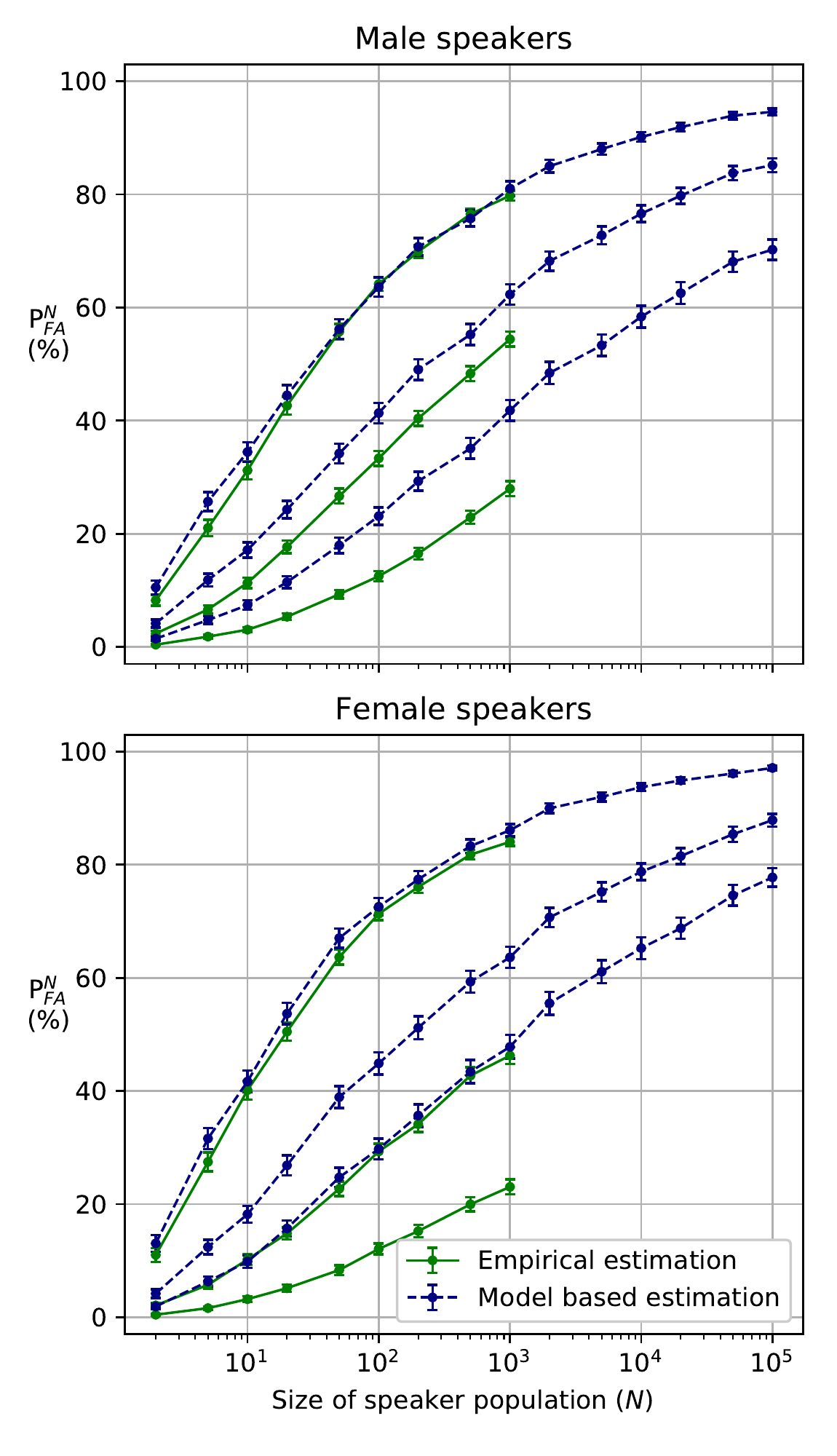}
    \caption{x-vector.}
    \label{fig:f2_2}
  \end{subfigure}
  \caption{Empirical and model-based estimates of worst-case false alarm rates with $N$ impostors} for various sizes of speaker populations. In each plot, three empirical and model-based estimates are shown for three different thresholds. These thresholds are obtained separately for each plot using cost parameters defined in Table \ref{tab:dcf_parameters}. The curves from top to bottom correspond to the thresholds of minDCF$_1$, minDCF$_2$, and minDCF$_3$, respectively. The model-based estimates follow closely emprical estimates for low threshold values, but as the threshold gets stricter, the difference between the empirical and model-based estimates grows. The curves are obtained using $T = 1000$ in Algorithms \ref{algo:fa-n-best} and \ref{algo:fa-n-best-model}. The mean values obtained using these algorithms are shown together with their 99\% confidence intervals. \label{fig:estimates}
\end{figure}

\begin{figure}[h]
  \begin{subfigure}[b]{0.5\textwidth}
    \includegraphics[width=\textwidth]{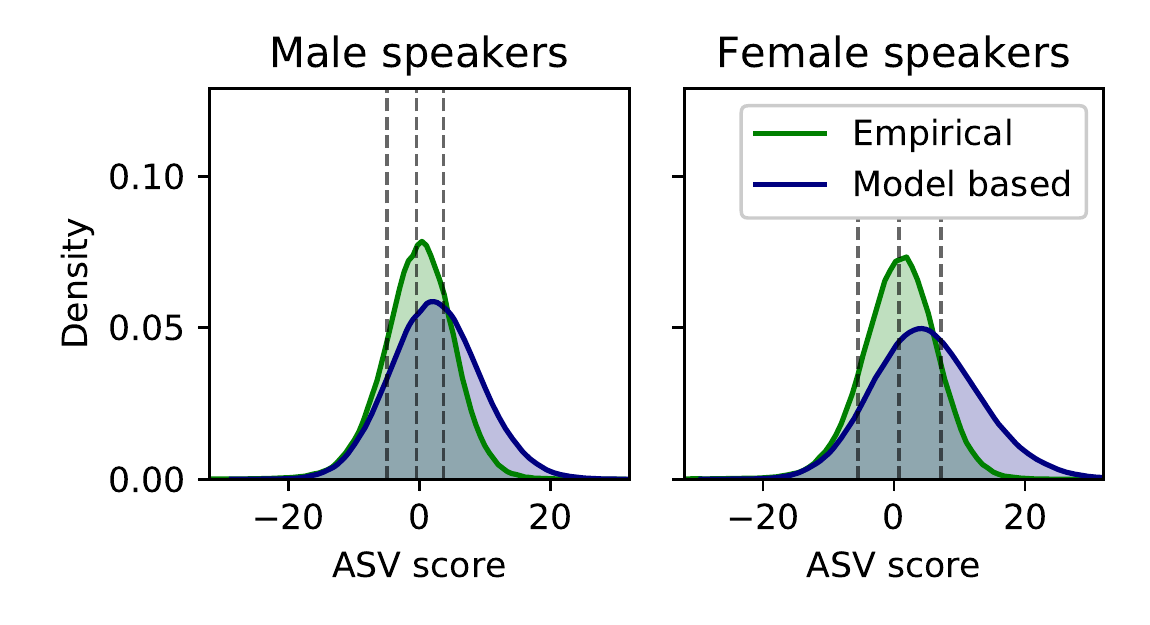}
    \caption{i-vector.}
    \label{fig:f1_3}
  \end{subfigure}
  \hfill
  \begin{subfigure}[b]{0.5\textwidth}
    \includegraphics[width=\textwidth]{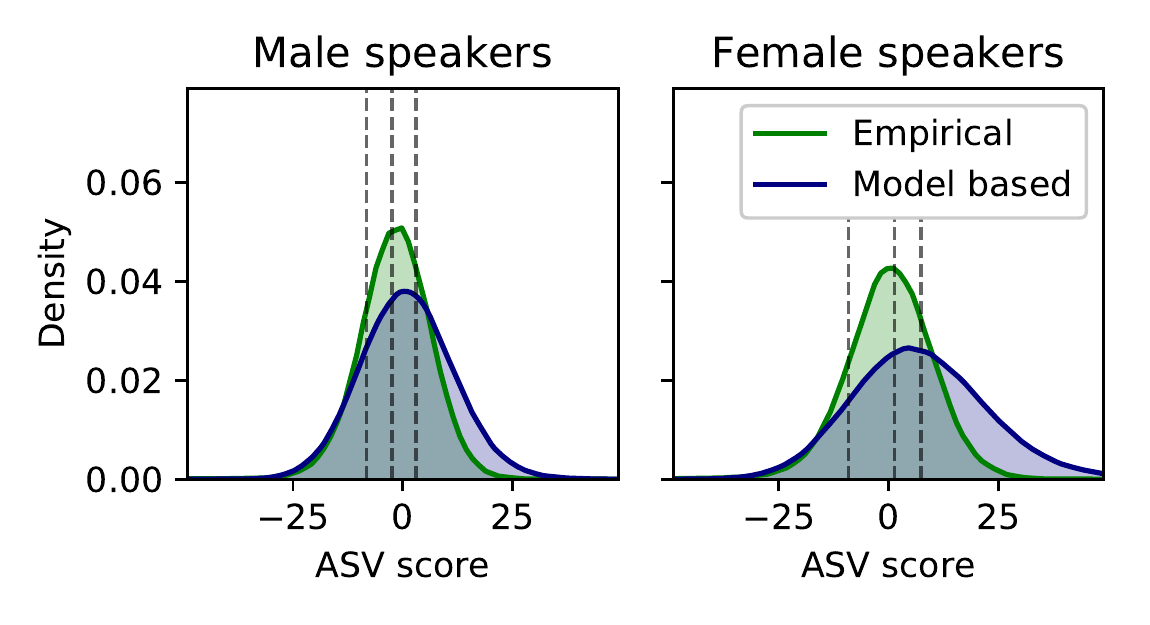}
    \caption{x-vector.}
    \label{fig:f2_3}
  \end{subfigure}
  \caption{Score distributions of the closest impostors ($N=1000$) pooled together from $T=1000$ samplings/simulations for the empirical and model-based approaches. Dashed lines represent minDCF thresholds obtained using different sets of cost parameters, which are presented in Table \ref{tab:asv-performance}. For the strictest threshold, the area under the density curves on the right side of the threshold is larger for the model-based estimation, which explains why the model-based estimation lead to higher false alarm rates as shown in Figure \ref{fig:estimates}. \label{fig:best_impostor_scores} }
\end{figure}

We estimated worst-case false alarm rates empirically using Algorithm \ref{algo:fa-n-best} by randomly selecting enrolled speaker $T=1000$ times. Similarly, we use $T=1000$ in Algorithm \ref{algo:fa-n-best-model} to obtain model-based estimates. The estimates are shown in Figure \ref{fig:estimates} for both ASV systems using three different thresholds obtained using the DCF parameter sets in Table \ref{tab:dcf_parameters}. We find that model-based approaches give good estimates when the threshold is low (higher cost for misses). When the threshold is higher than in the minDCF$_1$ case, the model-based estimates can be seen as a conservative upper bounds for the $\mathrm{P}^N_\text{FA}$ rates. We also find that the differences between the empirical and the model-based estimates are greater for females than for the males. To obtain further insight, we depict the score distributions of the closest impostors for population size of $N=1000$ in Figure \ref{fig:best_impostor_scores}. The figure indicates that especially for females, the model-based score distributions tend to be too wide and slightly shifted to the right, which causes higher false acceptance rates when a high threshold value is used.

Using the estimates, we can predict that for the minDCF$_1$ threshold, $\mathrm{P}^N_\text{FA}$ is \mbox{95 -- 98\%} for an impostor population of size $\num[group-separator={\text{\,}}]{100000}$. For minDCF$_3$ threshold, we can rely only on the empirical estimates, which tell us that, depending on the system and gender,  $\mathrm{P}^N_\text{FA}$ rate of \mbox{12 -- 28\%} is obtained for a population of size $1000$. Note that our populations contain only speakers from one gender. If the population would contain speakers from both genders, the false alarm rates would be lower.

\section{Discussion} \label{subsec:discussion}

Before concluding, the authors would like to address two relevant concerns, the over-estimated false alarm rates at high threshold, and the worst-case attack assumption.

\subsection{Analysis of the model-based worst-case false alarm estimation}\label{subsec:overestimation}

From Figure \ref{fig:best_impostor_scores}, we observe two apparent problems in the score distributions given by the model for the closest impostors:  a) they are shifted to the right and b) they have too large variances. As a result, some of the generated scores of the closest impostors are too high, which results in over-estimated false alarm rates given by the model. We have identified three causes for the problems.

\renewcommand\arraystretch{1.1}
\begin{table}[th]
\caption{Sources of mismatch between the observed and generated scores. The generative score model assumes that pairwise non-target scores and means of pair-wise scores ($\mu$) are normally distributed, while the analysis shows that they are skewed to the left. Additionally, scores with the closest impostors tend to have smaller variances than scores with random impostors, which is not factored into the model.}
\centering \small
\centerline{
\begin{tabular}{l C{1.5cm} C{1.5cm} C{1.5cm} C{1.5cm}}
\toprule
& \multicolumn{2}{c}{i-vector system} & \multicolumn{2}{c}{x-vector system} \\
%\cline{2-2}
& males & females & males & females \\
\midrule
Avg. skewness of pairwise scores & -0.20 & -0.29 & -0.20 & -0.27\\
Skewness of $\mu$ & -0.86 & -0.99 & -0.99 & -0.62\\
Avg. STDEV* of scores with the closest impostors & 5.1 & 5.5 & 7.5 & 9.4\\
Avg. STDEV* of scores with random impostors & 6.2 & 7.1 & 9.6 & 13.7\\
\bottomrule
\multicolumn{5}{r}{\footnotesize{* Computed as a square root of an average of variances.}}
\end{tabular}
}
\label{tab:discussion_analysis}
\end{table}
\renewcommand\arraystretch{1.0}

First, we found that the empirical distribution of $\mu$, which is the distribution of score means of speaker pairs, is skewed to the left (negative skewness, see Table \ref{tab:discussion_analysis}). Consequently, the fitted normal distribution (assumed in the model) has longer tail on the right than what the original score data had. The right tail of the distribution of $\mu$ is where we will find the closest impostors in Algorithm \ref{algo:fa-n-best-model}. As a result, the scores of the closest impostors are shifted to the right.

Additionally, we observed that the variation in target-vs-impostor scores is smaller for speaker pairs with the closest impostors than for random impostors. In other words, the closer the impostor's voice is to the target speaker's voice, the smaller is the variance in scores between the two speakers. As our model does not take this into account, the speaker pairs with closest impostors tend to have too large score variances.

Finally, the scores between the speaker pairs are also skewed to the left, which is another source of mismatch between empirical scores and scores generated by the model.

\begin{figure}[h]
\centering
\includegraphics[trim={0cm 9.7cm 0cm 0.8cm},clip, width=0.5\textwidth]{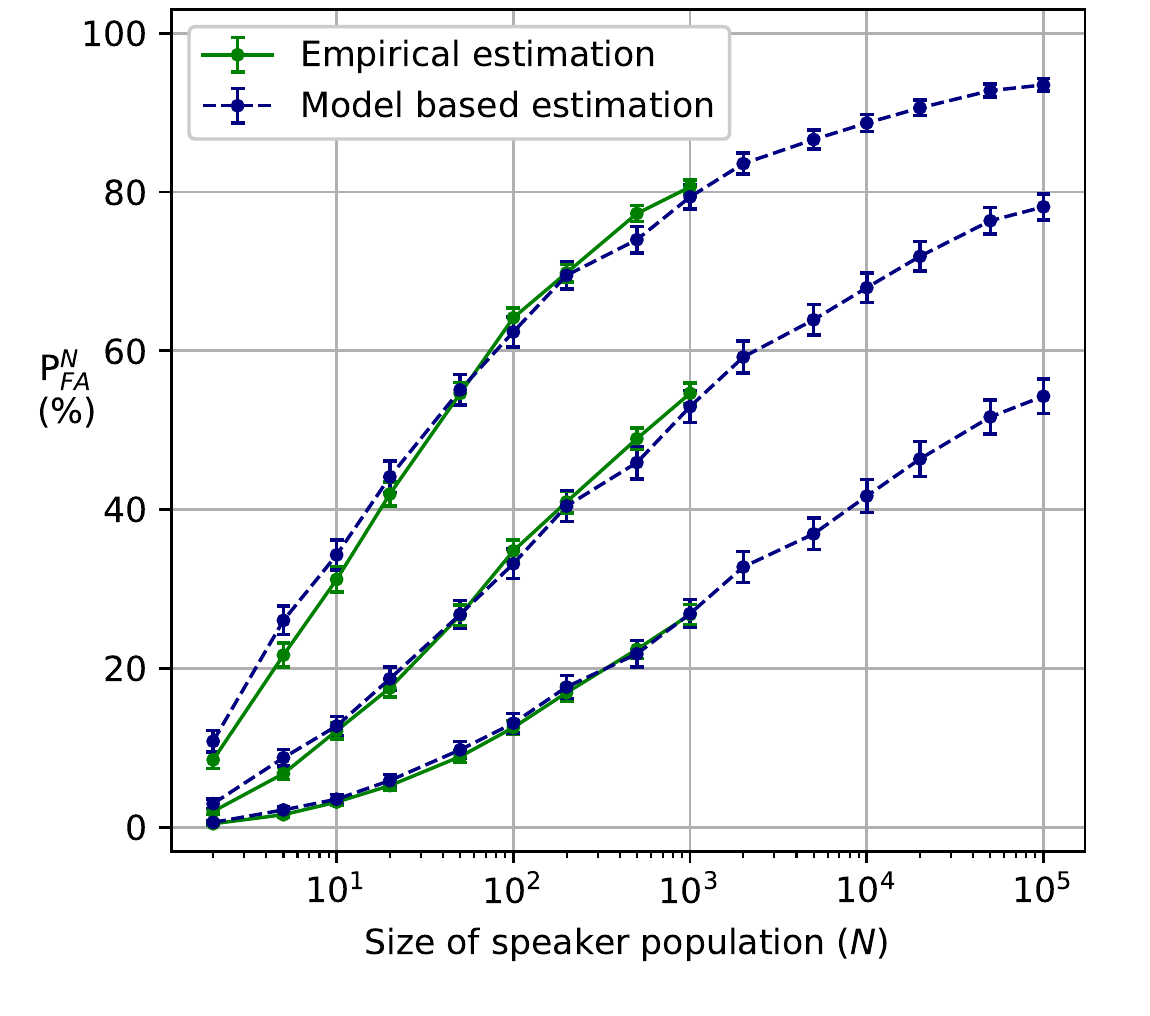}
\caption{Worst-case false alarm estimates for male scores given by x-vector system after tweaking the model hyper-parameters manually. First, the parameter $\alpha_\lambda$ was adjusted until the variances of distributions in Figure \ref{fig:best_impostor_scores} matched and then $\mu_0$ was adjusted to fix the shifting misalignment. As a result, the model-based estimates follow closely the empirical estimates unlike in Figure \ref{fig:estimates}.}
\label{fig:model_tweak}
\end{figure}

These observations open two potential directions towards increasing the prediction accuracy. The first direction is to revise the proposed generative model to take into account the skew of score distributions, as well as by relaxing the assumption of a shared variance. This can be done at the cost of losing conjugacy between distributions in the model, leading to increased computational complexity of hyper-parameter estimation. An alternative, second direction would be supervised fine-tuning to optimize some loss function between the empirical and the model-based estimates. As our model is parameterized only by six numbers, we believe that a good hyper-parameter configuration can be found in reasonable time using one of the \emph{derivative-free} optimization methods \cite{Larson-2019}. To give some empirical evidence for this claim, Fig. \ref{fig:model_tweak} displays an example where we tuned our model parameters \emph{manually}. As seen, the model itself is actually flexible enough to fit the empirical false alarm rates accurately --- but the purely generative training criterion does not find the parameter values that achieve this. With the manually corrected model, we obtained 54\% worst-case false acceptance rate estimate for population size of $\num[group-separator={\text{\,}}]{100,000}$ for the strictest minDCF$_3$ threshold, whereas the original model clearly over-estimated this by giving FA rate of 70\% as shown in the top-right panel of Figure \ref{fig:estimates}.

\subsection{False alarm estimation in simulated attack scenarios}

\begin{figure}[t]
  \begin{subfigure}[b]{0.5\textwidth}
    \includegraphics[width=\textwidth]{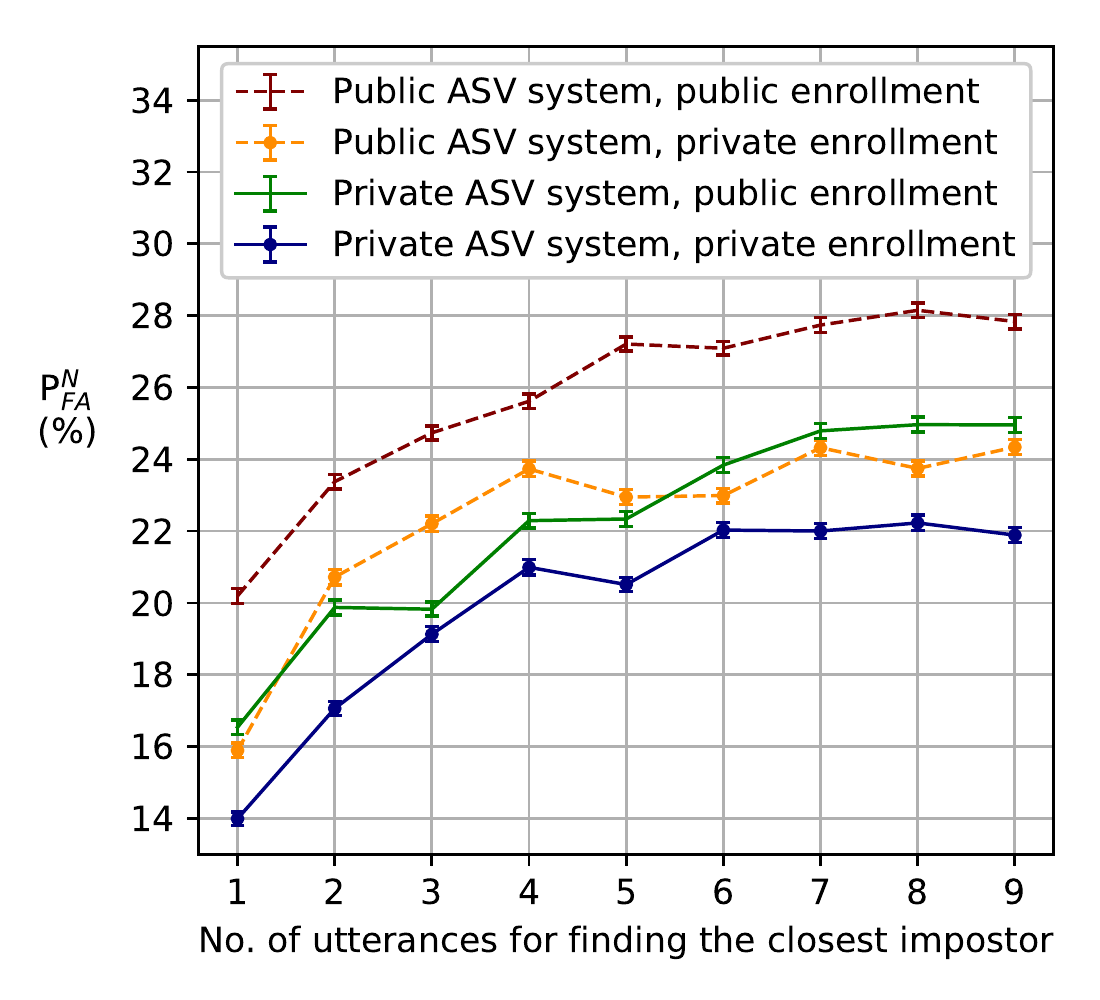}
    \caption{Male speakers.}
    \label{fig:f1_4}
  \end{subfigure}
  \hfill
  \begin{subfigure}[b]{0.5\textwidth}
    \includegraphics[width=\textwidth]{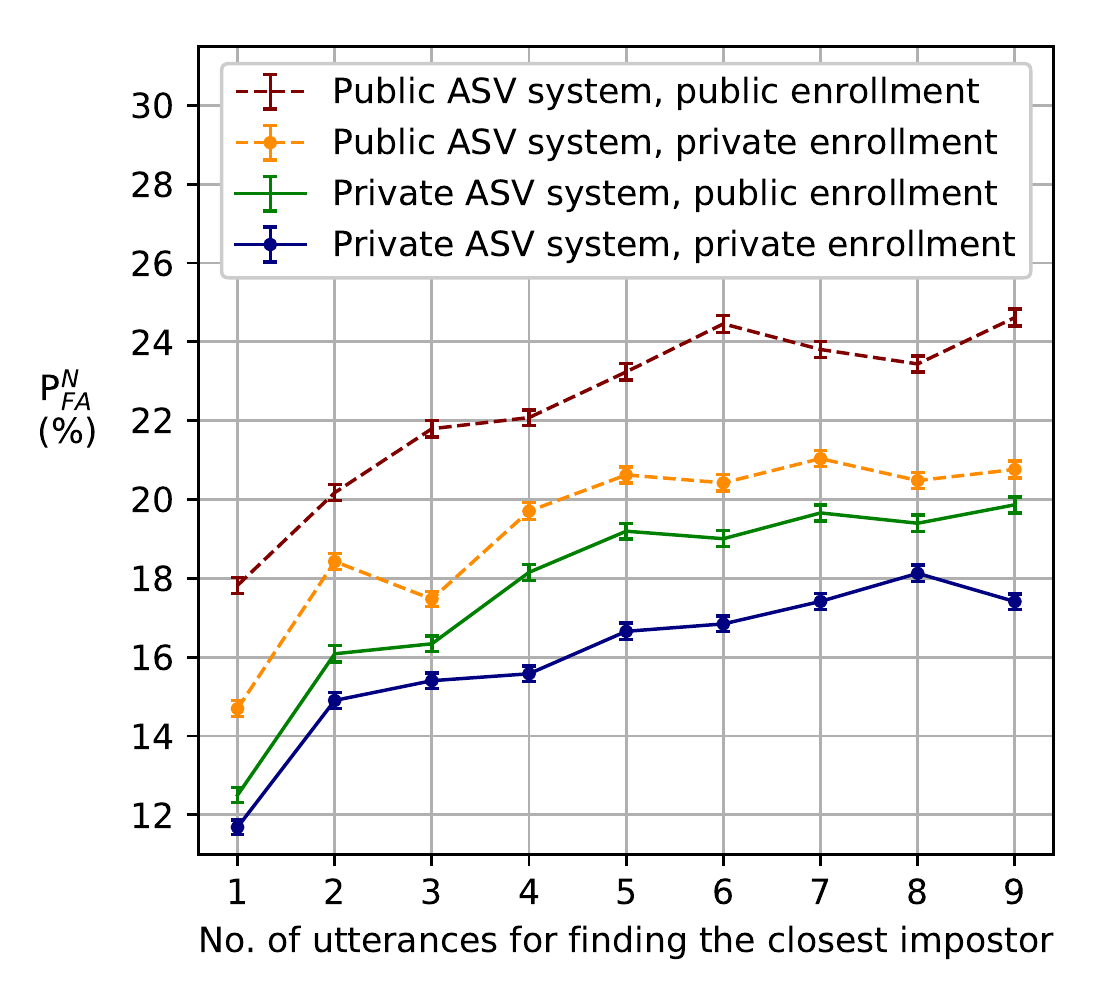}
    \caption{Female speakers.}
    \label{fig:f2_4}
  \end{subfigure}
  \caption{Empirical estimation of $\mathrm{P}^N_\text{FA}$ ($N=1000$) in various scenarios for x-vector system with minDCF$_3$ threshold (see Table \ref{tab:dcf_parameters}). `Public ASV system' refers to a case where the closest impostors in Algorithm \ref{algo:fa-n-best} are selected using the same x-vector system. To simulate a scenario, where attacker uses another ASV system for impostor selection due to not knowing the details of the deployed system (`private ASV system'), an i-vector system is used to select the closest impostors. Further, `public enrollment'  and `private enrollment' refer to such cases, where the attacker has access (public) or does not have access (private) to the enrolled target speaker's enrollment data. If the enrollment data and the ASV system are public, the selection of the closest impostor is easier, which results in higher false acceptance rates.} \label{fig:public_vs_private}
\end{figure}

So far we have considered worst-case false alarm estimation from the system deployer's perspective. From the presented results, we can gain understanding on how many enrolled speakers systems with specific thresholds can handle without starting to confuse speakers to each other too much.

Next, let us consider a scenario, in which a malicious attacker is utilizing ASV technology to find similar sounding speakers to the enrolled target speaker's voice to break the ASV system. As discussed in Section 1, the previously presented results can be considered as the worst-case situation, where the attacker has access to both the deployed ASV system as well as to the target speaker's enrollment data. In reality, the attacker would be unlikely to have access to either of them. Instead, the attacker would first have to set up another ASV system and then collect some speech data from the target speaker to perform the speaker search. These steps will make the attack more difficult as the closest impostor obtained using attacker's system and data might not be the same as what would be the closest impostor when using the attacked system and the real enrollment data.

To study the effect of system/data mismatch in the impostor selection, we set up a following experiment. First, we divided the available 18 utterances for each speaker into two disjoint sets of nine utterances. The first one was used for impostor selection and the second for speaker enrollment. We compared this setup to a case, where the same set of nine utterances was used both for impostor selection and enrollment to address the effect of data mismatch. Further, we also varied the number of utterances used for impostor selection from one to nine to see the effect of the amount of data used for impostor search. We simulated the ASV system mismatch by using i-vector system to select closest impostors, while the x-vector system was considered to be the attacked system. This was compared to the case where the impostor search was done using the same attacked x-vector system.

The results are shown in Figure \ref{fig:public_vs_private}, which reveals the expected patterns: when there is no data mismatch or ASV system mismatch, false acceptance rates are highest, which means that the attacks are most successful. If there is either data mismatch or system mismatch, the false acceptance rates drop. The lowest false acceptance rates are obtained, when both types of mismatches are present and when the number of utterances available for impostor search is low.

As another future direction, we consider designing a model for joint modeling of scores from two different ASV systems suitable for more realistic scenario where the attacker does \emph{not} have access to the target speaker's enrollment data and the deployed ASV system.

\section{Conclusions}

Seamless integration of artificial intelligence to our daily lives, including speech technology products, raises growing concern of their trustworthiness and safety. Our study resides in the landscape of automatic speaker verification (ASV), or voice biometrics security. One unique feature of voice (and face) biometrics is that, unlike traditional physical biometrics --- fingerprints, iris, retina, DNA to name a few --- is that much of the biometric data is \emph{publicly available in the Internet} through social media, news, interviews, lectures, and workplace websites to name a few. An important concern is the relation of false alarm (false acceptance) and database size: regardless of the selected ASV technology, given a large enough database, one will eventually have speaker collisions. A technology-aware attacker may increase the likelihood of such collisions through the use of public-domain ASV system  to identify target speakers from a public database \cite{Vestman2020-CSL-voice-mimicry}. Even without dedicated attacks, however, the number of different voices (subject to extrinsic and intrinsic speech variations) is not infinite. Therefore, eventually, ASV performance will be capped at some database size. 

The methodology concept put forward in this study gives us novel tool to address the dependency of false alarm rate and database size beyond the size of a given evaluation corpus. The proposed model produced reasonable match with empirical scores and displayed the expected trend of increasing false alarm rate as a function of database size. Our model is general and can be applied to analyze (and optimize) any black-box ASV system to produce graphs similar to those in Fig. \ref{fig:estimates}, based on detection scores only. As such, these graphs are \emph{predictions} by the model --- how the ASV system will behave if one were able to collect more speakers assuming the speaker sampling process remains the same. Even if it is not easy to experimentally validate the model beyond a given training corpus size, the general trends what we saw in our pilot experiments with VoxCeleb corpus are deemed as expected: false alarm rates increase as a function of database size and will eventually saturate. 

Our work has a number of limitations as well. First, as the results indicate, the generative model overestimated the false alarm rates, especially for the high-security operating region (high threshold). In real-world deployment of ASV, the detection threshold $\tau$ needs to be optimized to achieve a desirable security--convenience trade-off based on some development data and application (DCF setting). If the false alarm rate is over-estimated, the threshold $\tau$ would have to be increased (relative to the value it would have been set with precise knowledge of the false alarm rate), leading to decreased user convenience due to increased miss rate. Nonetheless, as Section \ref{subsec:discussion} indicates, our proposed generative model is flexible enough to be adjusted so that the empirical and predicted false alarm rates will match closely. Our model design philosophy has been simplicity: all the parameters are automatically learned from data and we leverage from conjugate families of distributions to enable efficient inference. The suggested future improvements include revising our distributional assumptions, and combining generative modeling with discriminative fine-tuning.

Second, our model assumes a worst-case scenario where the attacker has access to the target speaker's enrollment data, as well as the attacked ASV system. This is no different from standard NIST SRE style evaluations where an evaluator reports standard evaluation metrics (such as EER, minDCF, or $P_\text{FA}$) on a given, fixed evaluation corpus with known trial key. In future, we are interested in extending our generative model to model interaction between two different ASV systems and across different data domains.

Furthermore, it would be interesting to compare different ASV systems and database qualities. VoxCeleb data was selected for the experiments primarily due to the large number of speakers and the amount of intra-speaker scores. Nonetheless, being representative of \emph{found Internet data}, VoxCeleb contains many style, channel and environment variations. At least for academic curiosity, it would be interesting to repeat our simulations on  more controlled database for reference purposes. Further, it will be interesting to analyze the impacts of i-vector and x-vector dimensionality, dimensionality reduction of these embeddings, and speaker subspace size in PLDA. Finally, it would be interesting to apply the proposed methods to speaker diarization or other use cases within ASV, such as score normalization with increased speaker cohort size.

\section*{Acknowledgement}
The work has been supported by Academy of Finland (proj. no. 309629 entitled ``NOTCH: NOn-cooperaTive speaker CHaracterization'') and by the Doctoral Programme in Science, Technology and Computing (SCITECO) of the UEF. The work of the first author was partially financially supported by the Government of the Russian Federation (Grant 08-08). A part of the work of the third author was supported by NEC internship program. We gratefully acknowledge the support of NVIDIA Corporation with the donation of the Titan V GPU used for this research.

\section*{Appendix I: parameter inference}\label{sec:parameter-inference}

In the following we describe the algorithm used to estimate hyper-parameters of the proposed generative model. We find the values of hyper-parameters $\vec{\theta}$ that maximize the likelihood function -- 
the joint probability density of the observed data viewed as a function of $\vec{\theta} =\{\mu_0, \sigma_0^2, a_\sigma, b_\sigma, \alpha_\lambda, \beta_\lambda\}$:
\begin{gather}  \nonumber
\mathscr{L}(\vec{\theta}) = \int \prod_{i=1}^T \mathcal{N}(m_i|\mu_0, \sigma_0^2) \mathrm{Gam}(\lambda_i|\alpha_\lambda, \beta_\lambda) \mathrm{InvGam}(\sigma_i^2|a_\sigma, b_\sigma) \\  \nonumber
\prod_{j=1}^{N_i} \mathcal{N}(\mu_{i,j}|m_j, \lambda_i, \sigma_i^2) \prod_{l=1}^{L_{i,j}}  \mathcal{N}(s_{i,j,l}|\mu_{i,j}, \sigma_i^2) \dif m_i \dif \lambda_i \dif \sigma_i^2 \dif \mu_{i,j}
\end{gather}
For many probabilistic models with latent variables, including the proposed one, this objective function is intractable (cannot be evaluated). A commonly adopted strategy to avoid this obstacle is to use the \emph{expectation-maximization} (EM) algorithm \cite{Dempster-1977}, an iterative optimization method to find the local extrema of the likelihood function. The EM algorithm alternates between two steps: \textbf{expectation} step (E-step) and \textbf{maximization} step (M-step). On the E-step it computes (approximate) posterior distribution of the latent variables and on the M-step it updates all the hyper-parameters of the model. 

Inference for latent variable models can be conducted through \emph{variational Bayes} \cite[Chapter 10]{Bishop-book} or \emph{Monte-Carlo} techniques \cite[Chapter 11]{Bishop-book}. We choose the former approach due to its better scalability in terms of computational costs. The variational Bayesian inference approximates the exact posterior distribution $p(\{m_i\}, \{\lambda_i\}, \{\sigma_i^2\}, \{\mu_{i,j}\}|\{s_{i,j,l}\})$ by a \emph{variational} distribution $q$ from a restricted family of distributions. The variational distribution is found by minimizing the Kullback-Leibler divergence from the posterior distribution. 

One of the commonly adopted strategies, known as \emph{black-box variational inference} (BBVI) \cite{Ranganath-2014}, is to explicitly define the family of variational distributions and use stochastic optimization \cite{Robbins-1951} to minimize the objective.
Another strategy to define $q$ is the \emph{mean-field} approximation \cite{Bishop-book, Jordan-1999} which assumes the variational distribution to be fully factorized:
\begin{gather}
q(\{m_i\}, \{\lambda_i\}, \{\sigma_i^2\}, \{\mu_{i,j}\}) = \prod_{i=1}^T q(m_i) q(\lambda_i) q(\sigma_i^2) \prod_{j=1}^{N_i} q(\mu_{i,j}) 
\end{gather}
but with no further assumptions imposed on the functional forms of the factors. For conditionally conjugate models \cite{Wang-2013} this approach leads to closed form solutions in a coordinate descent optimization algorithm which iteratively updates the parameters of one factor while holding the others fixed. Since the proposed model is conditionally conjugate we choose to use the mean-field approach due to its lower computational complexity. The inference algorithm performs the following updates performed until convergence.

\noindent \textbf{\underline{Expectation step (E-step):}}
\vspace{1ex}
\begin{itemize}
\item Updating $q(m_i)$:
\begin{gather} \nonumber
q(m_i) = \mathcal{N} \left( m_i | \hat{m}_i, s_i^2  \right) \\ \nonumber
s_i^2 = \left( N_i \mathbb{E}\left[\lambda_i\right] \mathbb{E}\left[\frac{1}{\sigma_i^2}\right] + \frac{1}{\sigma_0^2} \right)^{-1} \\ \nonumber
\hat{m}_i = \left( N_i \mathbb{E}[\lambda_i] \mathbb{E}\left[\frac{1}{\sigma_i^2}\right] + \frac{1}{\sigma_0^2} \right)^{-1} \left( \mathbb{E}[\lambda_i] \mathbb{E}\left[\frac{1}{\sigma_i^2}\right] \left( \sum_{j=1}^{N_i} \mathbb{E}\left[\mu_{i,j}\right] \right) + \frac{\mu_0}{\sigma_0^2} \right) \\ \nonumber
\mathbb{E}[m_i] = \hat{m}_i, \ \mathbb{E}[m_i^2] = \hat{m}_i^2 + s_i^2
\end{gather}	

\item Updating $q(\sigma_i^2)$:
\begin{gather} \nonumber
q(\sigma_i^2) = \mathrm{InvGam} \left(\sigma_i^2 | \hat{a}_i , \hat{b}_i \right) \\ \nonumber
\hat{a}_i = a_\sigma + \frac{N_i}{2} + \sum_{j=1}^{N_i} L_{i,j} \\ \nonumber
\hat{b}_i = b_\sigma + \frac{1}{2} \mathbb{E}\left[ \sum_{j=1}^{N_i} \sum_{l=1}^{L_{i,j}} (s_{i,j,l} - \mu_{i,j})^2 \right] + \frac{1}{2} \mathbb{E}[\lambda_i] \mathbb{E}\left[ \sum_{j=1}^{N_i} (\mu_{i,j} - m_i)^2 \right] \\ \nonumber
\mathbb{E}\left[\frac{1}{\sigma_i^2}\right] = \frac{\hat{a}_i}{\hat{b}_i}, \ \mathbb{E}[\log \sigma_i^2] = \log \hat{b}_i - \psi(\hat{a}_i)
\end{gather}

\item Updating $q(\lambda_i)$:
\begin{gather} \nonumber
q(\lambda_i) = \mathrm{Gam} \left(\lambda_i | \hat{\alpha}_i , \hat{\beta}_i \right) \\ \nonumber
\hat{\alpha}_i = \alpha_\lambda + \frac{N_i}{2} \\ \nonumber
\hat{\beta}_i = \beta_\lambda + \frac{1}{2} \mathbb{E}\left[\frac{1}{\sigma_i^2}\right] \mathbb{E}\left[ \sum_{j=1}^{N_i} (\mu_{i,j} - m_i)^2 \right] \\ \nonumber
\mathbb{E}[\lambda_i] = \frac{\hat{\alpha}_i}{\hat{\beta}_i}, \ \mathbb{E}[\log \lambda_i] = \psi(\hat{\alpha}_i) - \log \hat{\beta}_i.
\end{gather}	

\item Updating $q(\mu_{i,j})$:
\begin{gather} \nonumber
q(\mu_{i,j}) = \mathcal{N} \left( \mu_{i,j} | \hat{\mu}_{i,j} , \hat{s}_{i,j}^2 \right) \\ \nonumber
\hat{\mu}_{i,j} = \frac{\sum_{l=1}^{L_{i,j}} + \mathbb{E}[\lambda_i]}{L_{i,j} + \mathbb{E}[\lambda_i]} \\ \nonumber
\hat{s}_{i,j}^2 = \left( \mathbb{E}\left[\frac{1}{\sigma_i^2}\right] (L_{i,j} + \mathbb{E}[\lambda_i]) \right)^{-1} \\ \nonumber
\mathbb{E}[\mu_{i,j}] = \hat{\mu}_{i,j}, \ \mathbb{E}[\mu_{i,j}^2] = \hat{\mu}_{i,j} + \hat{s}_{i,j}^2 
\end{gather}
\end{itemize}
Here, $\mathbb{E}[\cdot]$ denotes the expected value of a random variable.

\vspace{1ex}
\noindent \textbf{\underline{Maximization step (M-step):}}
\vspace{1ex}

Given an approximate posterior distribution found on the E-step, the M-step proceeds by updating the hyper-parameters $\vec{\theta}$ as follows:
\begin{itemize}
\item Updating $\mu_0$:
\begin{gather}  \nonumber
\mu_0 = \frac{1}{T} \sum_{i=1}^T \mathbb{E}[m_i] 
\end{gather}

\item Updating $\sigma_0^2$:
\begin{gather}  \nonumber
\sigma_0^2 = \frac{1}{T} \sum_{i=1}^T (\mathbb{E}[m_i] - \mu_0)^2
\end{gather}

\item Updating $\alpha_\lambda$ and $\beta_\lambda$:
\begin{gather}  \nonumber
\alpha_\lambda, \beta_\lambda = \arg \max_{\alpha, \beta} T(\alpha \log \beta - \log \Gamma(\alpha)) + (\alpha - 1) \sum_{i=1}^T \mathbb{E}[\log \lambda_i] - \beta \sum_{i=1}^T \mathbb{E}[\lambda_i]
\end{gather}

\item Updating $a_\sigma$ and $b_\sigma $:
\begin{gather}  \nonumber
a_\sigma, b_\sigma = \arg \max_{a, b} T(a \log b - \log \Gamma(a)) + (a - 1) \sum_{i=1}^T \mathbb{E}[\log \sigma_i^2] - b \sum_{i=1}^T \mathbb{E} \left[\frac{1}{\sigma_i^2} \right]
\end{gather}
\end{itemize}
The last two updates are two-dimensional convex optimization problems. Their solutions can be obtained using numerical optimization algorithms specialized to these tasks \cite{Minka-2002,Llera-2016}. Our approach is a less elaborate version of \cite{Minka-2002} where we use general-purpose root-finding algorithms which can be found in any commonly-adopted mathematical library.

The EM algorithm repeats the E- and M-steps outlined above until the convergence. In our experiments we found that a few iterations are sufficient to reach a point where any further iterations do not substantially change the values of hyper-parameters.

\section*{Appendix II: Score Distribution of the Model is Approximately Gaussian}\label{sec:scores-as-gaussian}

In the sequel we show how to obtain the marginal distribution of the observations
\begin{gather} \nonumber
p(s) = \int p(s|\mu, \sigma^2)p(\mu|m, \lambda, \sigma^2)p(m)p(\lambda)p(\sigma^2)  \dif \mu \dif m \dif \lambda \dif \sigma^2 
\end{gather}
by integrating out all the latent variables in the model one-by-one. We begin by noting that convolution of two Gaussians is another Gaussian with summed variances, to integrate out $\mu$:
\begin{gather} \nonumber
p(s)= \int \mathcal{N}(s|m, \sigma^2 + \sigma^2/\lambda) \mathcal{N}(m|\mu_0, \sigma^2_0) \mathrm{Gam}(\lambda|\alpha_\lambda, \beta_\lambda) \mathrm{InvGam}(\sigma^2|a_\sigma, b_\sigma) \dif m \dif \lambda \dif \sigma^2
\end{gather}
Further, since the inverse gamma distribution is a conjugate prior for Gaussian distribution with fixed mean, we arrive at the following:
\begin{gather} \nonumber
p(s) = \int t_{2a_\sigma}(s|m,b_\sigma/a_\sigma(1+1/\lambda)) \mathcal{N}(m|\mu_0, \sigma^2_0) \mathrm{Gam}(\lambda|\alpha_\lambda, \beta_\lambda) \dif m \dif \lambda  
\end{gather}
where $t_\nu(s|\eta, \varsigma^2)$ denotes the non-standardized $t$-distribution with $\nu$ degrees of freedom, mean $\eta$ and variance $\varsigma^2$. 
Since the $t$-distribution can be closely approximated by a Gaussian distribution, which is its limiting case when $\nu \to \infty$, even for moderate values of $\nu$, we can approximate the score distribution by a continuous mixture of Gaussians with gamma as the mixing distribution:
\begin{gather} \nonumber
p(s) \approx \int \mathcal{N}(s|\mu_0, \sigma^2_0 + b_\sigma/a_\sigma(1+1/\lambda)) \mathrm{Gam}(\lambda|\alpha_\lambda, \beta_\lambda) \dif \lambda  
\end{gather}
Note that the distributions inside the integral resemble a conjugate pair, which would lead to $p(s)$ being the $t$-distribution. Therefore, we speculate that the distribution $p(s)$ can be roughly approximated by a Gaussian. In fact, our simulations indicate that sampling scores from the model results in bell curve shaped histograms. The analysis above reveals a potential limitation of the proposed model -- the assumption that the distribution is symmetric around the mean.

% -------------------------------------------------------------------------
\bibliographystyle{elsarticle-num}

\end{document}